\documentclass[preprint,12pt,authoryear]{elsarticle}

\usepackage{amssymb}
\usepackage{rotating}
\usepackage{amsmath}
\usepackage[ruled]{algorithm}               % algorithm float environment
\usepackage{algpseudocode}                  % algorithmicx commands (\State,
\usepackage{tcolorbox}
\usepackage{tikz}
\usepackage{booktabs}
\usepackage[normalem]{ulem}

\usetikzlibrary{arrows.meta}

\journal{Journal of Computational Physics}

\begin{document}

\begin{frontmatter}



% \title{GPU-Accelerated Computing of Flow Physics around Moving and Deforming Bodies at Extreme-Scale using Sharp-Interface Immersed Boundary Method} %% Article title
\title{A Scalable Sharp-Interface Immersed Boundary Method for Moving Boundary Flow Simulation on Multi-GPU Systems} %% Article title

\author[label1,label2]{Sushrut Kumar}
\author[label2]{Joshua Romero}
\author[label2]{Massimiliano Fatica}
\author[label1,label2]{Jung-Hee Seo}
\author[label1]{Rajat Mittal}
\affiliation[label1]{organization={Johns Hopkins University},%Department and Organization
            % addressline={Department of Mechanical Engineering}, 
            city={Baltimore},
            postcode={21218}, 
            state={MD},
            country={USA}}

\affiliation[label2]{organization={NVIDIA Corporation},%Department and Organization
            city={Santa Clara},
            postcode={95051}, 
            state={CA},
            country={USA}}

%% Abstract
\begin{abstract}
Sharp-interface immersed boundary methods provide a versatile framework for simulating flows with complex moving boundaries, but their irregular data structures, dynamic sparsity, and discontinuous pressure operators present significant challenges for efficient multi-GPU implementation. This work presents a fully GPU-native, multi-GPU acceleration of the sharp-interface immersed boundary solver \texttt{ViCar3D}. The implementation combines a hybrid OpenACC/CUDA programming model with a matrix decomposition that isolates dynamically changing immersed-boundary operators, GPU-optimized geometric kernels, asynchronous dual-stream sparse matrix--vector execution, and a symmetric-memory-based sparse communication strategy. A GPU-oriented iterative solver employing a preconditioned stabilized bi-conjugate gradient (BiCGStab)  method is developed to efficiently treat the pressure Poisson equation on highly stretched Cartesian grids. The resulting framework preserves the accuracy and versatility of the original solver while delivering substantial acceleration and scalability for simulations involving complex moving and deforming immersed boundaries, thereby extending the applicability of sharp-interface immersed boundary methods to extreme-scale computational fluid dynamics.
\end{abstract}


% \begin{highlights}
% \end{highlights}

%% Keywords
\begin{keyword}

Scientific Computing \sep Computational Fluid Dynamics \sep Graphic Processing Units \sep Immersed Boundary Method
\end{keyword}

\end{frontmatter}


\section{Introduction}
\label{sec1}

Computational fluid dynamics (CFD) has become an indispensable tool for scientific discovery and engineering design, with applications spanning aerospace engineering, renewable energy, and biomedical flows \citep{stevens2017flow,taylor2009patient}. Increasingly, CFD is being applied to systems involving moving and deforming boundaries, fluid--structure interaction (FSI), and complex geometrical evolution, including bio-inspired propulsion, compliant structures, and physiological flows \citep{triantafyllou2000hydrodynamics,griffith2020immersed,verzicco2023immersed}. These applications often require high-fidelity simulations on geometrically complex domains while simultaneously demanding turnaround times compatible with design optimization, digital twins, and AI-enabled engineering workflows. Consequently, reducing time-to-solution through efficient utilization of modern GPU (Graphical Processing Unit) architectures has become a central challenge in computational fluid dynamics \citep{roccon2026gpu}.

For decades, body-fitted discretizations have been the dominant approach for maintaining geometric fidelity. However, their reliance on high-quality conforming meshes introduces significant overhead for moving-boundary problems, where repeated mesh deformation or regeneration increases computational cost, numerical complexity, and user effort \citep{mittal2005immersed,verzicco2023immersed}. Immersed boundary methods (IBMs) overcome these limitations by decoupling the representation of the immersed surface from the computational mesh, allowing the governing equations to be solved on a stationary Cartesian grid while imposing boundary conditions directly on the immersed geometry {\citep{Peskin1977-heart,mittal2005immersed,griffith2020immersed,mittal2023origin,verzicco2023immersed}}. 

Among the various IBM formulations, sharp-interface methods preserve the discontinuity of the solution at the immersed surface and achieve near body-fitted accuracy while retaining the simplicity of Cartesian meshes \citep{ye1999accurate,fadlun2000combined,mittal2008versatile,mittal2023origin,mittal2025freeman}.
Sharp-interface immersed boundary methods, however, present unique challenges for GPU implementation. Unlike conventional structured-grid solvers, moving-boundary sharp-interface methods continuously modify the immersed-boundary interpolation stencils, and boundary-adjacent control volumes. Consequently, the sparsity pattern of the discrete pressure operator changes every time step, producing \emph{dynamic sparsity}. This dynamic topology introduces irregular memory access, thread divergence, sparse communication, and continuously changing data structures that cannot be efficiently handled using conventional GPU sparse linear algebra techniques.

The present work focuses on the ghost-cell based sharp-interface solver \texttt{ViCar3D} \citep{mittal2008versatile}, which incorporates several features that present additional challenges for efficient GPU implementation. The solver enforces a Neumann pressure boundary condition on the immersed surface, introducing a discontinuous pressure Poisson operator that precludes FFT-based solvers and necessitates robust iterative methods. In addition, to maintain a high level of geometric and configurational flexibility for internal as well as external flows, \texttt{ViCar3D} supports highly stretched Cartesian grids and this coupled with combinations of Dirichlet, Neumann, and periodic external boundary conditions, places stringent demands on the pressure solver.

Although previous studies have demonstrated GPU implementations of sharp-interface immersed boundary methods \citep{viola2023gpu,viola2022fsei,raj2023gpu,vagnoli2026ib}, they do not address several unique challenges required for large-scale moving-boundary simulations on internal and external flows, including dynamic sparse operators, non-uniform Cartesian grids, discontinuous Neumann pressure operators, sparse communication, and scalable multi-GPU execution, that appear in the current class of sharp IBM solvers.

Building on our previous GPU acceleration of \texttt{ViCar3D} for stationary immersed boundaries \citep{kumar2026gpu}, this work presents a fully GPU-native, multi-GPU framework for large-scale simulations involving moving and deforming immersed bodies. The principal contributions include a dynamic sparse operator decomposition, GPU-optimized geometric classification kernels, an asynchronous dual-stream sparse matrix--vector execution strategy, an NVIDIA Shared Memory (NVSHMEM)-based sparse communication framework, and an asynchronous BiCGStab solver for the pressure Poisson equation. Together, these developments preserve the accuracy and versatility of the original sharp-interface immersed boundary method while enabling efficient and scalable execution on modern heterogeneous GPU architectures. 

The resulting framework enables high-fidelity fluid--structure interaction simulations at extreme scales, thereby advancing the applicability of sharp-interface immersed boundary methods to next-generation CFD workflows, digital-twin technologies, and large-scale scientific computing. Its accuracy, scalability, and computational performance are demonstrated through a series of benchmark and application cases presented in the following sections.

The developments presented here are relevant not only to the immersed-boundary community, but also to the broader CFD and computational mechanics communities because they address a general challenge: how to map dynamically changing, geometrically complex discretizations onto modern multi-GPU architectures without sacrificing accuracy or flexibility. The combination of dynamic operator decomposition, GPU-resident geometric processing, sparse communication, and asynchronous iterative solvers provides a framework that is potentially applicable to a wider class of moving-boundary and multiphysics simulations beyond the specific sharp-interface method considered here.

%% Use \subsection commands to start a subsection.
\section{\texttt{ViCar3D} -- A Sharp-Interface Immersed Boundary Solver}
\label{subsec1}
\subsection{Governing Equations and Numerical Method}
\begin{figure}[h!]
\centering
\includegraphics[width=1\textwidth]{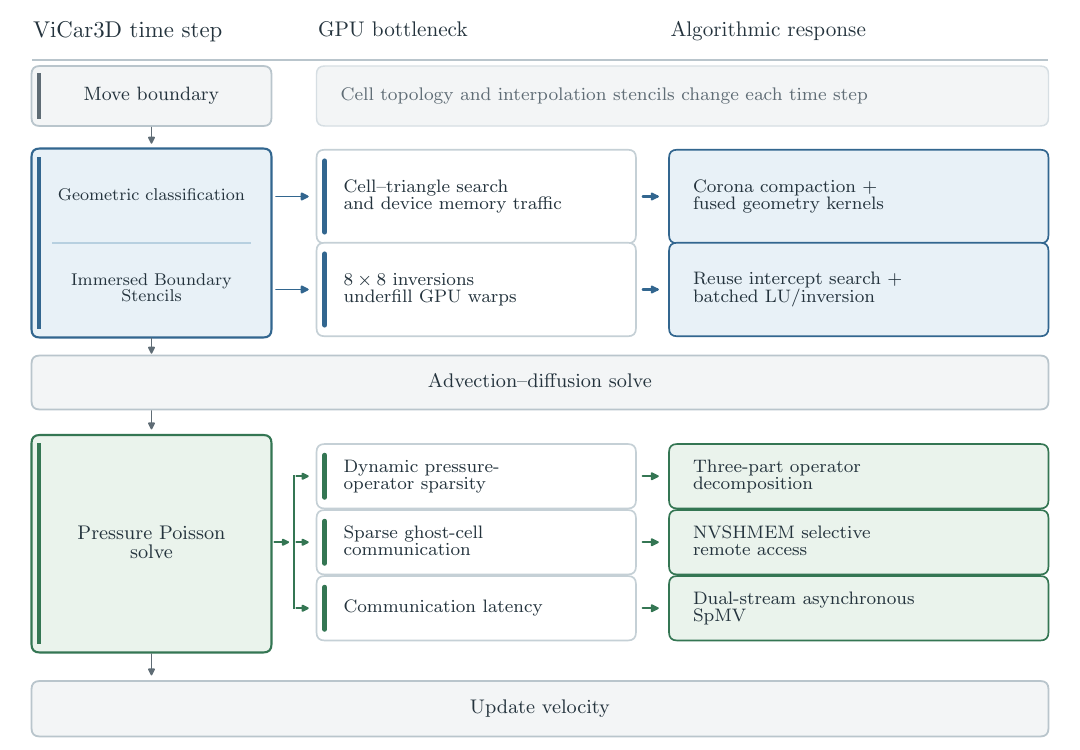}
\caption{\texttt{ViCar3D} time-stepping workflow showing the major computational bottlenecks arising from moving immersed boundaries, geometric processing, and the pressure Poisson solve, together with the GPU algorithms developed to address them.}
\label{fig:code_intro}
\end{figure}

The incompressible Navier--Stokes equations are solved on a non-uniform
Cartesian grid using the sharp-interface immersed boundary method of
\cite{mittal2008versatile,mittal2023origin}. Figure~\ref{fig:code_intro} summarizes the principal computational challenges and algorithmic innovations introduced in this work.

The time-stepping is initiated by moving the immersed body from $\mathbf{X}^{n+1} = \mathbf{X}^n + \Delta t \textbf{V}^{n+1/2 }$, where $\mathbf{X}$ is a vector prescribing the Lagrangian points that define the surface of the immersed body and $ \textbf{V}$ is the corresponding velocity of these points. Depending on the problem, this velocity might be prescribed or be the result of a fluid-structure interaction model. A fractional-step method is employed for time-stepping the flow in which an intermediate velocity field is first obtained from an advection--diffusion equation with explicit advection and implicit diffusion,
\begin{equation}
\frac{\mathbf{u}^*-\mathbf{u}^n}{\Delta t}
=
-\nabla\cdot(\mathbf{u^n \otimes u^n})
+\nu\nabla^2\mathbf{u}^*,
\label{eq:ade}
\end{equation}
with 
\begin{equation}
\mathbf{u} \left( \mathbf{X}^{n+1} \right)=\mathbf{V}^{n+1} \left( \mathbf{X}^{n+1} \right), 
\label{eq:adeBC}
\end{equation}
where $\mathbf{U}$ is the velocity on the immersed body surface. This is 
followed by the solution of the pressure via
\begin{equation}
\nabla^2p^{n+1}
=
\frac{1}{\Delta t}\nabla\cdot\mathbf{u}^* .
\label{eq:ppe}
\end{equation}
with \begin{equation}
\mathbf{n}.\nabla p^{n+1}\left( \mathbf{X}^{n+1} \right)=0.
\label{eq:ppeBC}
\end{equation}
The final step is the velocity update
\begin{equation}
\mathbf{u}^{n+1}
=
\mathbf{u}^*
-\Delta t\nabla p^{n+1} ,
\end{equation}
The application of the Neumann boundary condition for pressure is consistent with the fractional step method \citep{ye1999accurate} and ensures strict non-penetration into the immersed body surface at the end of the time-step.  This is key to ensuring a ``sharp'' representation in the IB method \citep{mittal2023origin}. Further details of the spatial and temporal discretization and the
sharp-interface formulation are provided in
\cite{mittal2008versatile,mittal2023origin}. 
%The pressure Poisson equation (PPE) represents the dominant computational cost of the flow solver and is therefore a primary focus of the GPU implementation developed here.

\subsection{Moving Immersed Boundaries: Computational Cost and Bottlenecks}
Immersed bodies are represented by unstructured triangulated surfaces. Boundary conditions (Eqs. \ref{eq:adeBC} and \ref{eq:ppeBC}) are imposed directly on the
physical surface using a ghost-cell formulation, thereby maintaining a sharp
representation of complex stationary, moving, and deforming geometries without
requiring body-conformal mesh generation \citep{mittal2008versatile}.

For moving or deforming bodies, the immersed-boundary geometry and its
relationship to the underlying Cartesian grid must be updated at every time
step. Cartesian cells are first classified as fluid or solid
(\texttt{IBlanking}), and ghost cells adjacent to the fluid region are then
identified. For each ghost cell, a boundary-intercept point and a corresponding
image point in the fluid are constructed, from which interpolation coefficients
are determined to impose the boundary conditions at the immersed surface. As
the body moves or deforms, the fluid--solid classification, ghost-cell
locations, and interpolation stencils change. Consequently, the
immersed-boundary contributions to the discrete pressure operator also evolve
in time. Efficiently accommodating this dynamically changing sparsity pattern
is a distinctive and significant challenge in mapping the sharp-interface
method to GPUs.

\begin{figure}[h!]
\centering
\includegraphics[width=1\textwidth]{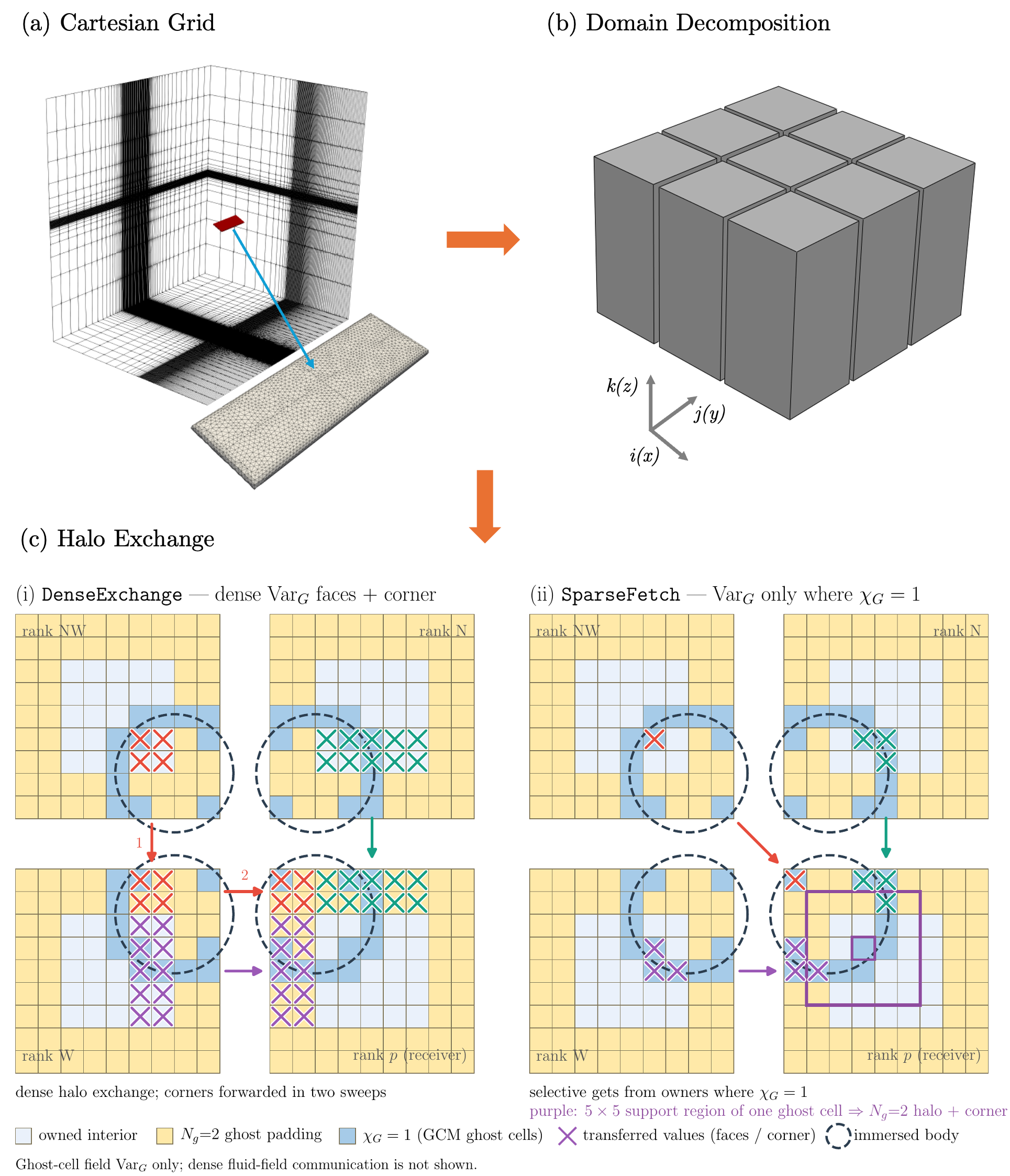}
\caption{Multi-GPU domain decomposition and communication. (a) Immersed body in a Cartesian grid. (b) Partitioning in the $x$--$y$ plane. (c) Ghost-field communication using dense \texttt{DenseExchange} (i) and selective \texttt{SparseFetch} (ii).}
\label{fig:domain_decomp}
\end{figure}

The legacy \texttt{ViCar3D} flow solver of \cite{mittal2008versatile} is implemented in Fortran and parallelized using MPI with a two-dimensional pencil decomposition in the $x$- and $y$-directions (Fig.~\ref{fig:domain_decomp}b). Three kernels are particularly important for GPU acceleration: geometric classification (\texttt{IBlanking}), the advection--diffusion equation (ADE) solve, and the pressure Poisson equation (PPE) solve. \texttt{IBlanking} classifies fluid and solid points and constructs the ghost-cell geometry and interpolation stencils, with its cost increasing with grid resolution and immersed-surface complexity. The ADE advances the intermediate velocity field using an alternating-direction line GS--SOR procedure; although its cost scales linearly with the number of grid points, line-wise data dependencies constrain parallelism. The PPE is solved using a preconditioned stabilized bi-conjugate gradient (BiCGStab) method. Its iterative sparse-matrix operations and dynamically changing immersed-boundary contributions make it the dominant computational kernel.

\begin{table}[h!]
    \centering
    \caption{Wall-clock time per time step for the three dominant kernels in
    the legacy MPI-based CPU implementation on the Rockfish cluster. Each
    compute node contained two 24-core Intel Xeon Gold 6248R processors. The
    smallest case used one processor socket, corresponding to half of a
    compute node.}
    \label{tab:cpu_component_timing}
    \begin{tabular}{lccccc}
        \hline
        Grid ($N_x \times N_y \times N_z$)
        & Nodes
        & CPU sockets
        & \texttt{IBlanking} (s)
        & ADE (s)
        & PPE (s) \\
        \hline
        $256 \times 320 \times 320$
        & $1/2$ & 1 & 0.700 & 0.951 & 21.234 \\

        $410 \times 510 \times 510$
        & 2 & 4 & 1.260 & 1.045 & 21.244 \\

        $910 \times 1110 \times 1110$
        & 5 & 10 & 12.200 & 2.623 & 55.701 \\
        \hline
    \end{tabular}
\end{table}

To establish a CPU performance baseline, the three dominant computational kernels were profiled for three-dimensional flow over a finite-aspect-ratio rectangular wing adapted from \cite{menon2022contribution} (Fig.~\ref{fig:domain_decomp}a). Table~\ref{tab:cpu_component_timing} reports the corresponding wall-clock time per time step over a range of grid sizes. For grids containing $\mathcal{O}(10^7)$--$\mathcal{O}(10^8)$ points, the PPE dominates the computational cost, accounting for approximately 90\% of the combined time spent in the three profiled kernels, while ADE and \texttt{IBlanking} contribute substantially less. For the largest case, containing approximately one billion grid points, the PPE requires 55.7~s per time step, while \texttt{IBlanking} and ADE require 12.2~s and 2.6~s, respectively. Notably, the fraction of the combined kernel cost associated with \texttt{IBlanking} increases from approximately 3\% for the smallest case to 17\% for the largest case. Thus, although the PPE remains the primary computational bottleneck, immersed-boundary geometric operations become an increasingly significant fraction of the computational cost as the problem size grows. This trend provides additional motivation for the GPU algorithms developed in the following sections, which target not only acceleration of the PPE but also efficient treatment of the evolving immersed-boundary geometry.

\section{GPU-Acceleration Implementation}
\subsection{Programming Model and Software Architecture}
To address the computational bottlenecks identified in the previous section and fully exploit the massive parallelism of modern Single Instruction, Multiple Thread (SIMT) GPU architectures, the legacy solver was extensively refactored into a hybrid GPU programming framework. The foundational implementation combines MPI for distributed-memory parallelism with OpenACC \citep{openacc34}. OpenACC provides a portable, directive-based programming model for offloading the majority of the computational workload while maintaining a single source code across CPU and GPU platforms. However, the most performance-critical kernels require finer control over thread hierarchy, memory layout, and kernel execution than can be achieved using directive-based programming alone. Many of these kernels were therefore reprogrammed using CUDA Fortran \citep{ruetsch2024cuda}, enabling direct access to the underlying GPU execution model and allowing architecture-specific optimizations. 

Efficient interoperability between the OpenACC runtime and CUDA Fortran is achieved through the \texttt{host\_data use\_device} directive, which exposes device memory pointers directly to CUDA kernels. This zero-copy interface eliminates unnecessary host-device memory transfers and enables both programming models to operate on the same device-resident data structures. The framework further leverages highly optimized vendor libraries through standard language bindings. Specifically, the cuBLAS Fortran API \citep{nvidia_cublas} is employed for dense linear algebra operations. We also utilize the C++ Thrust library \citep{bell2012thrust} to perform parallel operations such as prefix-sum required for array compaction entirely on the GPU. This hybrid software architecture combines the portability and rapid development afforded by OpenACC with the fine-grained control and high performance of CUDA Fortran, while seamlessly integrating optimized GPU libraries into the overall solver infrastructure.

\subsection{Multi-GPU Domain Decomposition and Communication Architecture}
\label{sec:multi_gpu}
%---------------------------------------------------------------

High-fidelity simulations of complex phenomena such as multiscale
fluid--structure interactions require large spatial and temporal resolutions,
leading to substantial computational and memory demands. While modern GPUs
provide increasing on-chip compute capability through large numbers of CUDA
cores, the available memory on a single device remains
limited for very large simulations. Consequently, distributing the
computational domain across multiple GPUs is not merely advantageous but often
essential for achieving the required problem size and performance. The
distributed-memory framework used for this purpose is based on the Message
Passing Interface (MPI) standard \citep{gropp2014using,mpi41standard}, which
provides a portable mechanism for process management and inter-process
communication across a wide range of high-performance computing architectures.

In our implementation, we partition our 3-D computational domain defined as $\Omega=[0,L_x]\times[0,L_y]\times[0,L_z]$ into an MPI-managed pencil-like topology in the $x$--$y$ plane as shown in figure~\ref{fig:domain_decomp}.
Given a process grid of $M_p^x\times M_p^y=P$ ranks, the global cell
counts are distributed such that the local cell count in the $x$
direction for rank $r$ is
\begin{equation}
N_x^{(r)}
=
\left\lfloor \frac{N_x^G}{M_p^x} \right\rfloor
+
\begin{cases}
1, & c_x^{(r)} < N_x^G \bmod M_p^x,\\
0, & \text{otherwise}.
\end{cases}
\label{eq:decomp}
\end{equation}
where $c_x^{(r)}=0,\ldots,M_p^x-1$ is the rank's Cartesian coordinate in the $x$ direction, with an analogous expression for $N_y^{(r)}$. This algorithm distributes remainder cells over the lowest Cartesian coordinates, ensuring that local extents differ by at most one cell
in each partitioned direction. While this load-balances the fluid grid cell-counts, the immersed-boundary workload may still vary significantly between ranks. We implement a specialized sparse matrix-vector multiplication (SpMV) optimized to efficiently handle this imbalance, which will be discussed in the section \ref{sec:solver}. 

The solution to governing equation require numerical approximation to derivatives near subdomain boundaries. The Ghost-Cell Method (GCM) introduces a geometric challenge as its multidimensional interpolation stencil can extend across the diagonal corners where the four MPI ranks meet. We make use of two sets of data points: one that resides in halo layers and the other, the ghost cells from the neighboring regions (see figure \ref{fig:domain_decomp}). Halo layers are padded grid points ($N_{\mathrm{gl}}$) around the partitioned domain in the $x$ and $y$ directions whereas Ghost cells are points within the solid body with neighboring fluid points (further details in the next section). The flow properties from these points can be received from neighboring rank using two mechanisms: a dense halo exchange and a sparse halo exchange. The dense halo exchange, termed here as \texttt{DenseExchange}, moves the entire slab of data whereas the sparse exchange (``\texttt{SparseFetch}''), selectively fetch data based on an indicator field function. 

For a scalar field, the leading-order communication volume is $\mathcal{O}(N_{\mathrm{gl}}(N_xN_z+N_yN_z))$ values per rank, where $N_x$, $N_y$, and $N_z$ denote local cell counts.
This dense exchange is executed either through GPU-aware MPI or through the NVIDIA Collective Communication Library (NCCL)~\citep{nvidia_nccl}. The NCCL backend supports CUDA streams and graph capture, allowing communication to be integrated into the solver's execution schedule. In the dense backend, both the fluid and ghost-cell components of the pressure-solver vectors are exchanged through complete halo buffers. While dense halo exchange is appropriate for the fluid component, applying the same approach to the ghost-cell field (referred to here as $p_G$) transfers many inactive entries. The ghost-cell distribution is sparse and irregular, with only a fraction of the halo region containing values required by the GCM.

To address this inefficiency, an alternative communication backend uses NVSHMEM \citep{nvidia_nvshmem}, NVIDIA's OpenSHMEM-based library for GPU clusters, which provides a partitioned global address space (PGAS). NVSHMEM is initialized using MPI, and remote owners are identified from the Cartesian domain decomposition. Device-initiated, one-sided remote memory access operations allow each GPU to fetch values directly from neighboring GPUs, including diagonal neighbors (see figure \ref{fig:domain_decomp}). Thus, corner values are obtained from their owners without forwarding through an intermediate rank. In this backend, the fluid component is fetched densely, whereas the ghost-cell component is fetched only where the mask satisfies $\chi_G=1$. In the NVSHMEM backend, \texttt{DenseExchange} retains dense transfers for the regular fluid component. \texttt{SparseFetch} retrieves remote ghost-field values only where $\chi_G=1$, with $\chi_G$ marking active immersed-boundary ghost cells (Figure~\ref{fig:domain_decomp}(c.ii)). The same operations apply to the intermediate Krylov vectors, with $p_G$ denoting the ghost-field component in the schematic. If $N_{G,\mathrm{halo}}$ denotes the number of active ghost-cell
entries in the remote halo, the ghost-cell payload scales as $\mathcal{O}(N_{G,\mathrm{halo}})$ rather than the complete halo volume because inactive ghost-cell entries are omitted. 

The next subsection describes the geometric classification that determines the ghost-cell mask used by \texttt{SparseFetch} along with GPU acceleration of the Ghost-cell methodology. The pressure solver then schedules the \texttt{DenseExchange} and, in the sparse variant, the \texttt{SparseFetch} concurrently with independent interior computations, completing communication before the dependent boundary operations (Section~\ref{sec:solver}).

\subsection{Geometric State Classification}
\label{sec:blanking}
\begin{figure}[htbp]
\centering
\includegraphics[width=0.9\textwidth]{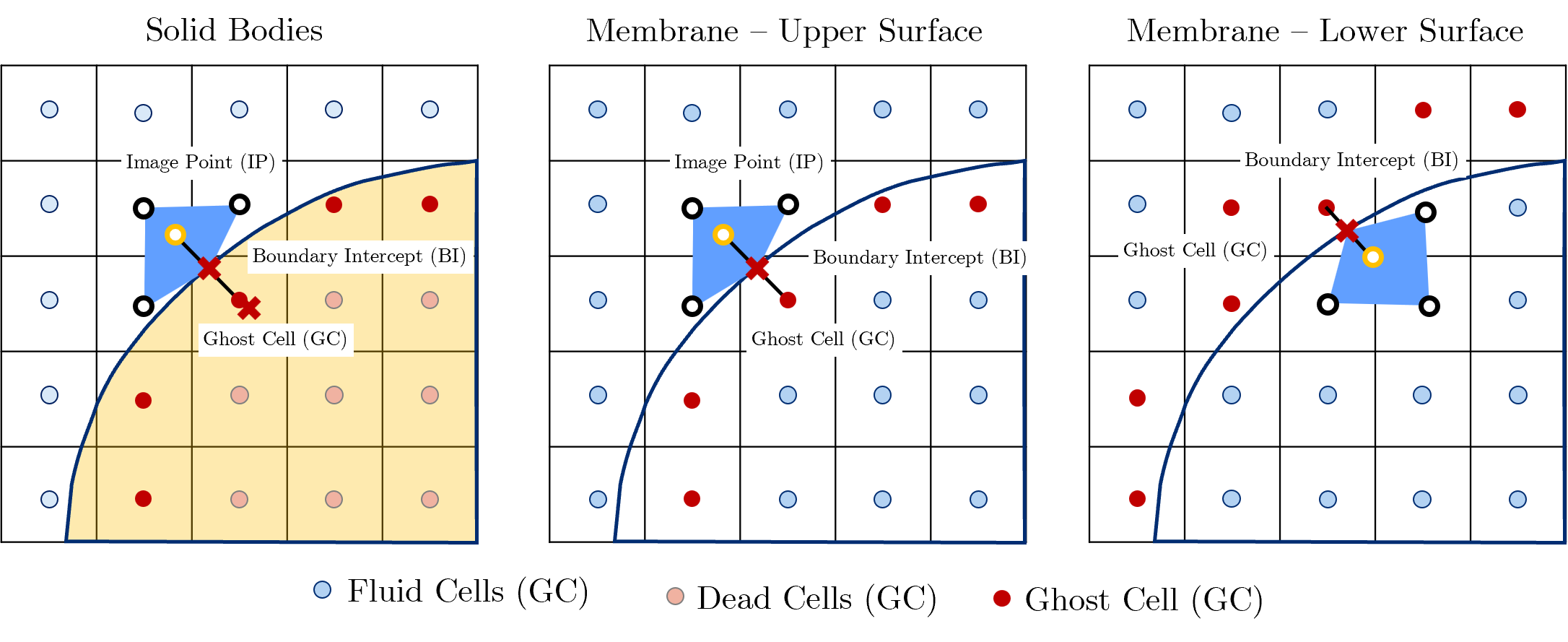}
\caption{Ghost-cell and image-point constructions for a volumetric solid (left) and a zero-thickness shell (middle and right).}
\label{fig:membrane_topology}
\end{figure}

The Ghost-Cell Method (GCM) imposes immersed-boundary conditions through interpolation along the local body normal. For each ghost cell ($\mathbf{x}_\text{GC}$), or fresh cell ($\mathbf{x}_\text{FC}$) generated by body motion, the method determines the corresponding boundary-intercept point ($\mathbf{x}_\text{BI}$) and an image point ($\mathbf{x}_\text{IP}$) in the fluid:
\begin{equation}
\mathbf{x}_\text{IP}
= \mathbf{x}_\text{GC}
+ L_{\text{probe}}
(\mathbf{x}_\text{BI}-\mathbf{x}_\text{GC}),
\label{eq:imagepoint}
\end{equation}
where $L_{\text{probe}}$ is the probe length. Construction of these interpolation stencils therefore requires identification of cells adjacent to the immersed surface and determination of their associated boundary intercepts. Moreover, we design the process so it can work for solid bodies and shells (zero-thickness bodies) as well as for configuration which may involve multiple bodies of both types. We define this process of classification as \texttt{IBlanking}. For volumetric solid bodies, this process classifies Cartesian grid cells as fluid or solid and identifies ghost cells as cells in the solid that have at least one immediate neighbor in the fluid.  All other cells in the solid interior are blanked, leaving a single layer of ghost cells along the fluid--solid interface (Figure~\ref{fig:membrane_topology}, left). Zero-thickness shells, such as membrane wings \citep{kumar2025computational}, fish fins \citep{kumar2026flow}, and flags \citep{orrego2017harvesting}, present a more complex topology because active fluid exists on both sides of the immersed surface. Therefore, shells require independent ghost-cell and image-point constructions on the two sides to preserve discontinuities across the shell as shown in figure~\ref{fig:membrane_topology}. We employ a dual-field approach using two variables, \texttt{var} and \texttt{var}$_G$ to handle all these body topologies and to easily program the iterative solvers as described in later sections. For moving and deforming bodies, the geometric state of affected Cartesian cells changes continuously, requiring \texttt{IBlanking}, ghost-cell identification, and boundary-intercept construction to be updated at every time step. Efficient geometric classification on GPUs is therefore a critical component of the moving-body algorithm. 

\begin{figure}[htbp]
 \centering
 \includegraphics[width=1\textwidth, height=0.8\textheight, keepaspectratio]{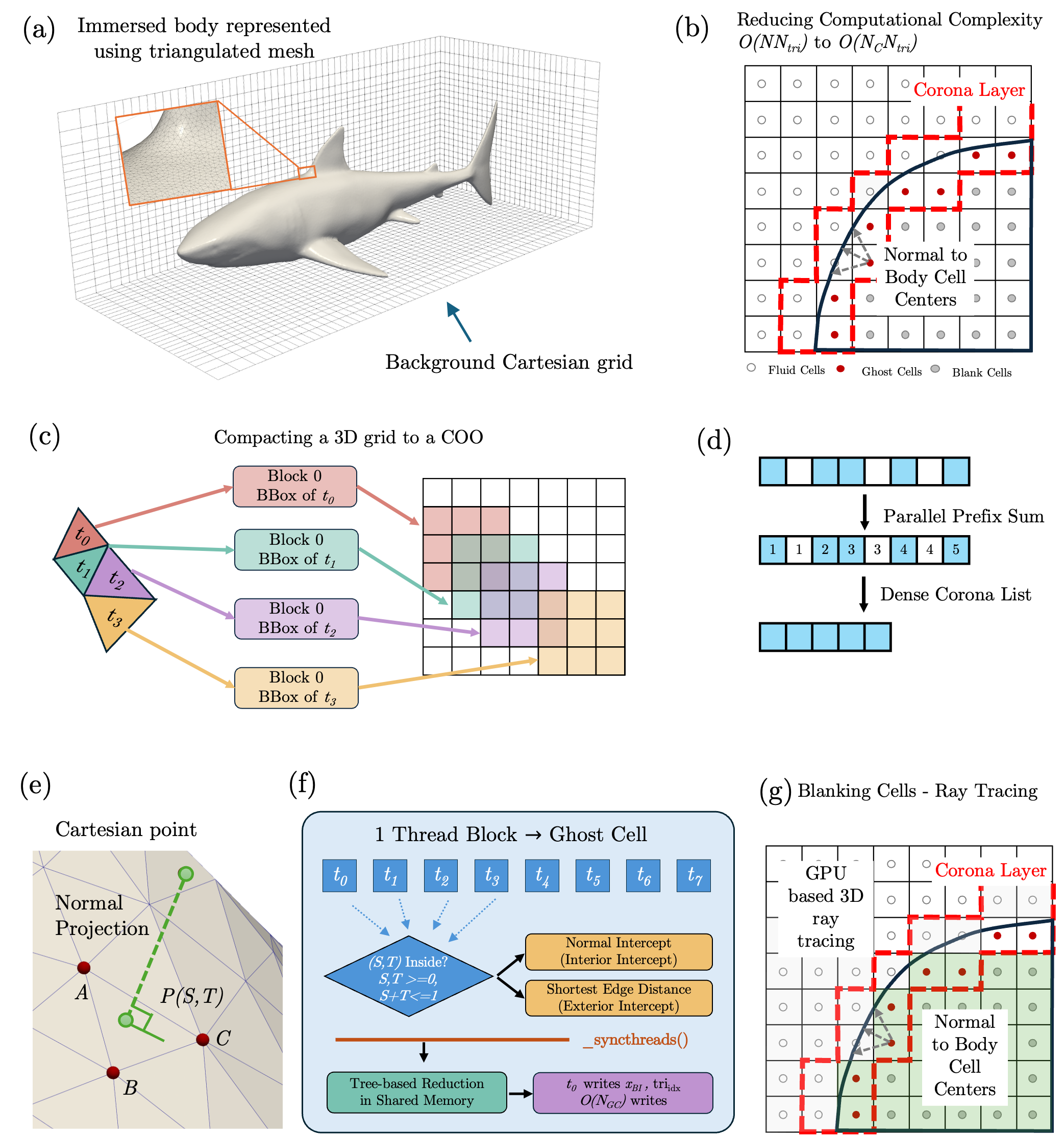}
\caption{GPU geometric preprocessing. (a) Surface mesh and Cartesian grid. (b) Corona layer and cell classification. (c) Corona identification and compaction. (d) Prefix sum for array compaction for ghost cell data. (e) Local inside/outside test. (f) Boundary-intercept search. (g) Ray-tracing fill.}
\label{fig:blanking_schematic}
\end{figure}

A direct geometric-classification strategy would determine the surface triangle nearest to each Eulerian cell and classify the cell from the orientation of the displacement vector relative to the corresponding surface normal. For a GPU subdomain containing $N=N_xN_yN_z$ Eulerian cells and an immersed surface containing $N_{\Delta}$ triangles, an exhaustive nearest-triangle search scales as $\mathcal{O}(N N_{\Delta})$. A naive GPU implementation further compounds this cost by decomposing the operation into separate kernels for distance evaluation, storage of cell-triangle distances, nearest-neighbor reduction or sorting, and final geometric classification. Explicitly storing these distances would require $\mathcal{O}(N N_{\Delta})$ global-memory traffic and storage. Such an implementation can create performance bottleneck for the large Cartesian grids and complex immersed geometries considered here. Consequently, an efficient GPU implementation must avoid both the exhaustive cell-triangle search and the global memory traffic of huge intermediary data. We alleviate these inefficiencies using three highly specialized kernel fusing many intermediary operations. The kernels work on corona identification, vertex selection, and projection (Figure~\ref{fig:blanking_schematic}) and only output the final data to device memory.

\paragraph{Phase 1: Body Corona Cell Identification}
The first kernel launch acts as a spatial filter and generates work for the next two phases of the geometric process. Here, we start with the grid setup as shown in figure \ref{fig:blanking_schematic}(a) and search a local ``corona'' region near the immersed body surface as shown figure \ref{fig:blanking_schematic}(b). A binning kernel assigns one thread block per surface element as shown in figure \ref{fig:blanking_schematic}(c). Threads locate the Cartesian cell-index intervals bounding the element and extend them by the user-defined \texttt{coronaSize} margin. This identifies a corona of $N_c$ candidate cells around the surface. The resultant array is a 3D point cloud around the body surface which scales with $N_c \propto (N^{2/3} \times$\texttt{coronaSize})$ \, \ll N$, and which is then subjected to a rigorous evaluation by the \texttt{iBlanking} routine. Working with this highly sparse non-contiguous memory access data is extremely inefficient. To alleviate this, we use a parallel prefix sum and scatter to compact the binary corona mask into a dense, contiguous list of Cartesian cell indices in global memory as shown in figure~\ref{fig:blanking_schematic}(d). Subsequent geometric-search kernels operate on this list and this reduces the overall complexity from $\mathcal{O}(NN_{\Delta})$ to $\mathcal{O}(N_c N_{\Delta})$ with $N_c \ll N$.

\paragraph{Phase 2: Nearest-Vertex Identification}
A rigorous evaluation of \texttt{IBlanking} for each corona cell requires the identification of the nearest vertex to each cell as shown in figure \ref{fig:blanking_schematic}(e). Here, the second kernel launches $N_c$ independent thread block, each assigned to a candidate corona cell, executing concurrently on GPU. Let $B$ be the CTA size and $N_m$ the number of surface vertices for the body being processed. Threads process vertices in strides and retain the top-$K$ nearest candidates using shared-memory bitonic sorting. Retaining the top-$K$ candidates, rather than only the nearest triangle-vertex, ensures robustness in geometrically ambiguous regions such as concave corners, where multiple triangles may admit valid normal projections and the closest boundary intercept is not necessarily associated with the nearest vertex. Bitonic sort is selected because its deterministic execution pattern eliminates warp divergence while requiring a critical-path depth of only
$
 D_{\text{bitonic}}(B)=\mathcal{O}\!\left(\log^2 B\right),
 \label{eq:bitonic_complexity}
$
corresponding to the number of sequential synchronization stages within each thread block. Over $\lceil N_m/B\rceil$ strides, for fixed $K$, the overall execution span is therefore
$T_{\mathrm{Phase2}}=\mathcal{O}\!\left(\lceil N_m/B\rceil\log^2 B\right)$,
allowing the globally nearest candidate vertex to be identified with logarithmic synchronization overhead while keeping all computation within shared memory.

\paragraph{Phase 3: Fused Analytical Projection and Reduction}
In this step, we identify whether a corona cell lies inside or outside a solid body. Here we assign a thread block for each corona cell and threads process candidate triangles incident to any of the $K$ retained vertices and execute exact analytical projections. For each candidate triangle, the barycentric coordinates $(S,T)$ (see figure \ref{fig:blanking_schematic}(e)) of the normal projection from the corona cell to the triangle determine whether the normal projection lies within the triangle. A projection is interior when $S\geq0$, $T\geq0$, and $S+T\leq1$. Refer to appendix \ref{app:barycentric} for mathematical formulation. %We encounter certain edge cases where barycentric coordinate based classification fails and we use the shortest distance to the triangle edges there. 
Figure~\ref{fig:blanking_schematic}(f) illustrates this search when the kernel is applied to corona cells. The projection tests and distance calculations are fused in one kernel. Geometry-dependent branches introduce warp divergence but avoid intermediate global-memory round-trips. This is acceptable as global-memory data writes incur a higher cost than warp-divergence. A shared-memory tree reduction selects the nearest valid candidate in $\mathcal{O}(\log B)$ stages, after which one thread writes the winning triangle index.

At the completion of these three phases we have a rigorous determination of the geometric state (inside? or outside?) of each corona cell. We then use parallelized ray-tracing procedure to mark all the other Cartesian cells in the domain as shown in figure \ref{fig:blanking_schematic}(g).

\subsection{Local Boundary Condition Enforcement and Batched Matrix Inversion}
Once the Eulerian cell classification is completed, the identification of Ghost Cells is simply done by identifying solid cells that have one immediate neighbor along any of the Cartesian directions that is a fluid cell. For shells, Ghost Cells are all boundary adjacent cells that have a neighbor on the other side of the shell. The next step is the construction of the interpolation scheme which requires the normal intercept from each ghost cell to the boundary. For this, we repeat the boundary-intercept search for each ghost cell as described earlier using the same geometric kernels. We then compute the boundary intercepts by normal projection (Figure~\ref{fig:blanking_schematic}(e)), followed by image-point construction using Eq.~\eqref{eq:imagepoint}. Once the normal projections successfully define the Image Point ($\mathbf{x}_\text{IP}$) within the fluid, the solver must construct a localized trilinear interpolation stencil to impose the specific boundary conditions (see \cite{mittal2008versatile} for mathematical details). For each generic scalar field $\phi$ evaluated at the 8 surrounding Cartesian nodes, the interpolation at the image point is defined as $\phi_\text{IP} = \sum_{i=1}^{8} \beta_i \phi_i$. The polynomial coefficients $\{C\}$ are derived by inverting an $8 \times 8$ local Vandermonde matrix $[V]$ mapping the physical coordinates:
\begin{equation}
 \{C\} = [V]^{-1}\{\phi\}, \quad \text{where} \quad \beta_i = f(C, \mathbf{x}_\text{IP}).
\end{equation}
With the weights $\beta_i$ established, the ghost cell value ($\phi_\text{GC}$) is computed analytically to strictly satisfy the boundary conditions at the boundary intercept. For Dirichlet conditions, this yields:
\begin{equation}
 \phi_\text{GC} = 2\phi_\text{BI} - \sum_{i=1}^{8} \beta_i \phi_i
\end{equation}
For Neumann conditions, the physical probe length is
$\ell_\text{probe}
=L_\text{probe}\|\mathbf{x}_\text{BI}-\mathbf{x}_\text{GC}\|$.
With the normal directed from the ghost cell toward the image point,
\begin{equation}
\phi_\text{GC}
= -\ell_\text{probe}
\left(\frac{\partial\phi}{\partial n}\right)_\text{BI}
+\sum_{i=1}^{8}\beta_i\phi_i.
\end{equation}
From a GPU computing perspective, independently inverting $N_\text{GC}$ independent $8 \times 8$ Vandermonde matrices ($N_\text{GC}$ is the number of ghost cells) presents a significant optimization hurdle. The individual matrices are far too small to saturate a CUDA warp, but too numerous to process sequentially. We resolve this by utilizing \texttt{cuBLAS} batched subroutines (\texttt{cublasDgetrfBatched} for LU factorization and \texttt{cublasDgetriBatched} for matrix inversion). This allows the hardware to invert thousands of local stencils concurrently to maximize throughput.

As the body (both solid and shell) undergoes large-scale deformations or movement through the fluid grid governed by discrete time updates (${\mathbf{X}^{n+1} = \mathbf{X}^n} + \mathbf{V}^{n+1/2} {\Delta t} $), the topological state of the Eulerian cells continuously changes. As illustrated in Figure \ref{fig:blanking_schematic} (Bottom Right), cells that were previously ghost cells frequently emerge into the active fluid domain, becoming "Fresh Cells." Because these newly active cells lack valid pressure and velocity histories from the previous time-step, their governing equations cannot be integrated normally. 

Before the linear solver is initialized, these Fresh Cells again undergo an identical local geometric identification procedure as Ghost cells established previously to establish a physics-consistent initial state. Leveraging the same \texttt{cuBLAS} batched Vandermonde inversion framework ($[V] \in \mathbb{R}^{8 \times 8 \times N_F}$), the fresh cell scalar value ($\phi_F$) is populated directly from the surrounding active fluid:

\begin{equation}
 \phi_F = \sum_{i=1}^{8} \beta_i \phi_i.
\end{equation}

%-------------------------------------------

\subsection{Iterative Solvers and Preconditioning Strategy}
\label{sec:solver}

The fractional-step formulation requires solves for the intermediate velocity through the Advection--Diffusion Equation (ADE) (eq. \ref{eq:ade}) and for pressure through the Pressure Poisson Equation (PPE) (equation \ref{eq:ppe}). The strongly diagonally dominant ADE system is efficiently solved by Jacobi iteration. The PPE lacks the strong diagonal contribution from the time derivative and grid stretching along with immersed-boundary coupling further deteriorate its conditioning. In literature, Incomplete LU (ILU) factorization and algebraic multigrid (AMG) are commonly used preconditioners for pressure Poisson systems. Their setup depends on the matrix connectivity and coefficients, which complicates their use when immersed-body motion changes the operator. As immersed bodies move through the Cartesian grid, the fluid-solid classification changes continuously, modifying the matrix connectivity at every time step. Changes in connectivity can require updates to the ILU factorization structure or the AMG hierarchy. Repeated setup introduces graph traversal and sparse-data restructuring costs. Moreover, the forward and backward triangular solves inherent to ILU have data dependencies that can limit available parallelism during preconditioner application.

To circumvent these issues, we have employed a Jacobi preconditioned BiCGStab solver for solving the PPE system. With $K=\operatorname{diag}(A)$, each preconditioner application is a pointwise multiplication by the inverse diagonal. The ghost-cell equations have unit diagonal, so their corresponding preconditioner entries are unity. These operations can be fused with the Krylov vector updates and require neither symbolic factorization nor triangular solves. This choice provides low setup cost and an embarrassingly parallel preconditioning on the GPU.

\begin{sidewaysfigure}[p]
\centering
\includegraphics[width=1\textheight]{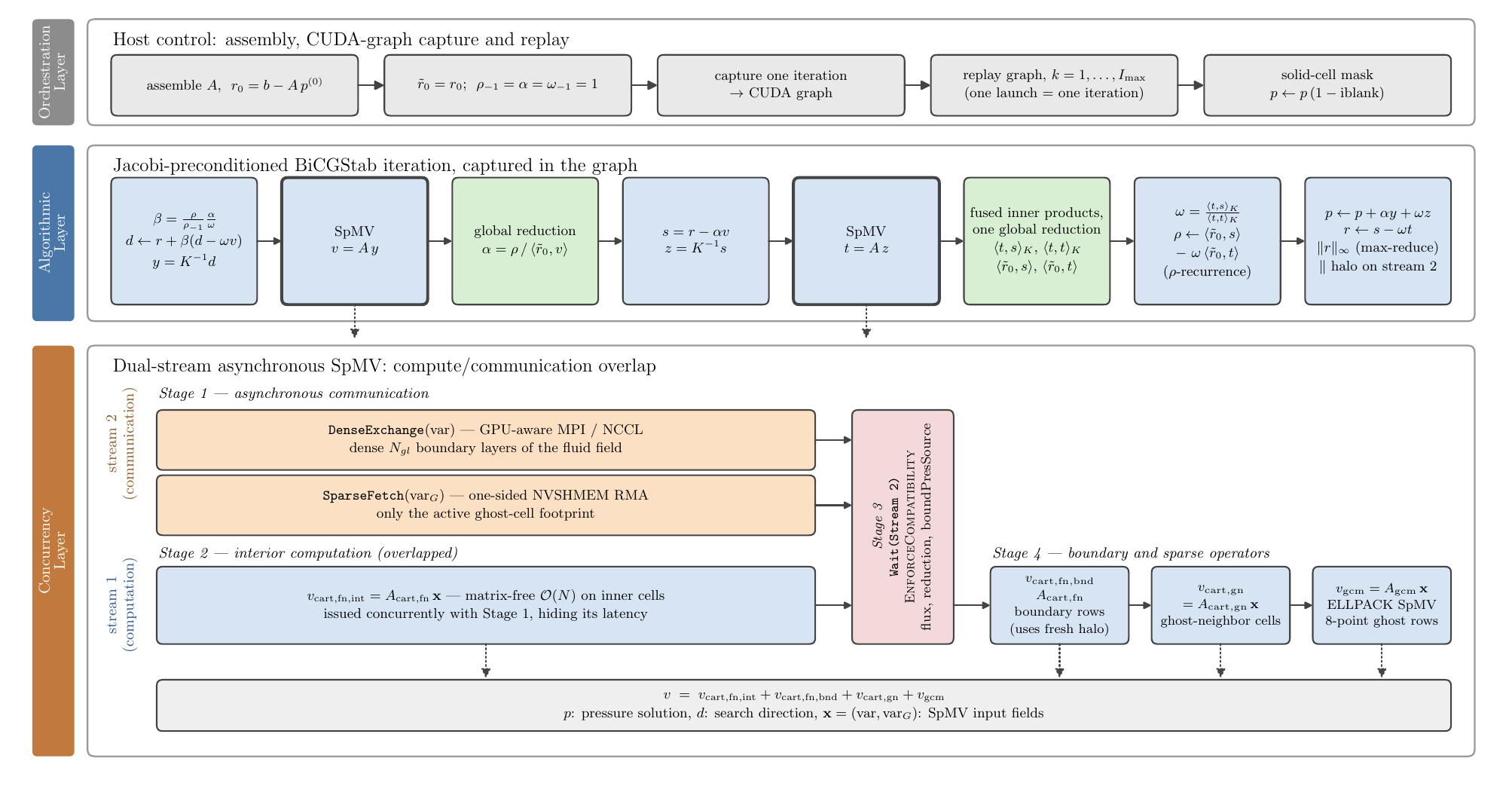}
\caption{GPU pressure-Poisson solve: host orchestration (top), Jacobi-preconditioned BiCGStab iteration (middle), and dual-stream matrix--vector product with communication overlap (bottom).}
\label{fig:bicgstab}
\end{sidewaysfigure}

As discussed earlier, body motion continuously changes the sparsity pattern of the discrete operator. Reconstructing and reallocating a monolithic sparse matrix at every time step would be particularly expensive on GPUs. However, the Cartesian grid method provides an advantage in dealing with dynamic sparsity of operator. Here we can easily isolate the dynamic immersed-boundary contributions from the regular Cartesian operator. This is shown in figure~\ref{fig:bicgstab} where the contributions to the pressure operator can be decomposed as
\begin{equation}
A\mathbf{x} =
\underbrace{A_{\text{cart}}\mathbf{x}}_{\text{Cartesian cells}}
+
\underbrace{A_{\text{gcm}}\mathbf{x}}_{\text{Ghost cells}}.
\label{eq:Asplit}
\end{equation}

The dominant Cartesian operator, $A_{\text{cart}}$, retains the regular
7-point stencil and is evaluated matrix-free using coalesced memory access.
The operator $A_{\text{gcm}}$ represents the GCM interpolation equations
associated with the $N_{\text{GC}}$ active ghost cells. Because each
ghost-cell row contains at most eight interpolation coefficients, the
coefficients and corresponding stencil indices are stored compactly in
ELLPACK format~\citep{ZHENG20142639}, while the unit diagonal associated
with the ghost-cell unknown is applied separately. A key GPU challenge
arises for fluid cells adjacent to ghost cells. In these locations, the
standard Cartesian stencil coefficient must be replaced to account for the
implicitly imposed boundary condition. Since the BiCGStab sparse
matrix-vector products are memory-bound, their performance is limited
primarily by data movement to and from device memory rather than floating-point
throughput. It is therefore inefficient to fetch additional
ghost-cell-related data for every fluid cell, since most fluid cells are not
adjacent to a ghost cell and would incur this memory traffic unnecessarily.

\begin{figure}[htbp]
\centering
\includegraphics[width=\textwidth]{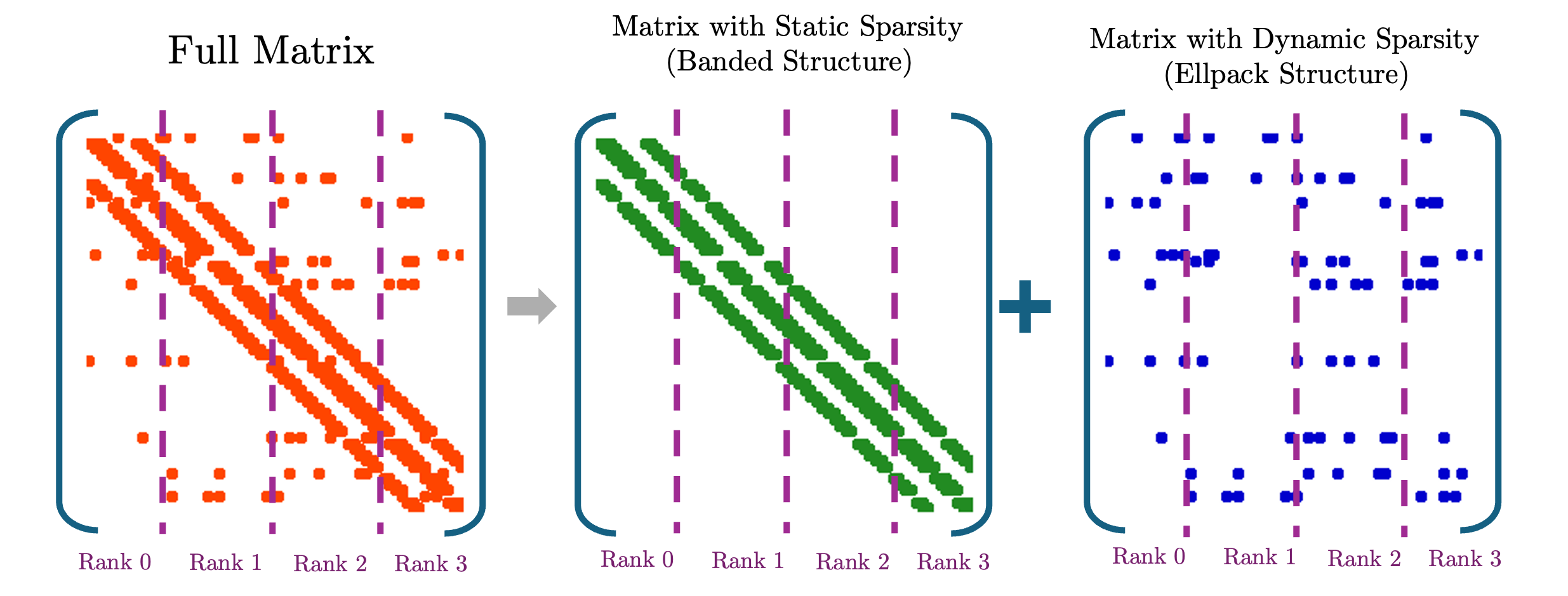}
\caption{Splitting of sparse operator for efficient implementation on GPUs}
\label{fig:matrix_split}
\end{figure}
We alleviate this inefficiency by further splitting the operator into ones with neighboring fluid cells or ghost cells, $A_{\text{cart}}\mathbf{x} = A_{\text{cart,fn}}\mathbf{x} + A_{\text{cart,gn}}\mathbf{x}$. The $A_{\text{cart,gn}}$ operator contains fluid cells whose Cartesian stencil includes ghost-cell values and selectively accesses the regular or ghost pressure fields according to the local boundary topology. The $A_{\text{cart,fn}}$ only contains cells with all neighboring cells are fluid cells.
This decomposition updates the ghost-cell interpolation rows and the masks identifying ghost-neighbor cells in vicinity of the body. The regular Cartesian stencil retains fixed connectivity and matrix-free evaluation. The resulting SpMV has arithmetic complexity
\begin{equation}
\mathcal{O}(N+8N_\text{GC}), \qquad N_\text{GC} \ll N,
\end{equation}
thereby retaining the efficiency of the structured Cartesian discretization while accommodating the evolving immersed-boundary topology. Details of the ghost-neighbor blending, ghost-cell interpolation stencils, and GPU data layout are provided in~\ref{app:matrix_operators}.

The custom BiCGStab formulation to invert the decomposed PPE matrix efficiently across multiple GPUs using the dual-field approach is summarized in figure \ref{fig:bicgstab}. Our implementation structure follows three layers: orchestration layer, algorithmic layer and concurrency layer. 

The top layer of Figure~\ref{fig:bicgstab} summarizes the host orchestration of the solve. One BiCGStab iteration is captured as a CUDA graph and replayed during the solve, reducing repeated host launch overhead. The middle algorithmic layer shows the multiple SpMV operations and the intervening vector updates along with global reductions. In this layer, multiple inner products are fused and their collective operations are grouped to reduce kernel launch overheads. During the execution, the coefficient computation introduces global value summation (also known as reduction) before the dependent Krylov updates. These global reductions use NCCL in both the dense-communication and NVSHMEM sparse-communication variants. The lower layer shows how dense and sparse communication is overlapped with computation during SpMV per iteration. The source and prescribed normal fluxes must satisfy the global Poisson compatibility condition due to presence of Neumann boundary conditions. Here we need to compute a global source term which has to be leaked out of all the outward faces of the computational domain. The solver enforces compatibility through a global flux reduction and applies the distributed boundary-source correction stored in \texttt{boundPresSource}. These flux calculations require updated neighboring values. The compute stream therefore waits for communication to finish before applying the compatibility correction and evaluating the remaining boundary contributions. 

The dual-stream pipeline follows four stages (Figure~\ref{fig:bicgstab}, bottom). First, Stream 2 exchanges neighboring data. The regular Cartesian field requires halo communication and in \texttt{DenseExchange}, the fluid and ghost field is exchanged through dense buffers. In \texttt{SparseFetch}, NVSHMEM fetches only the remote ghost-cell values required by the local stencils. Second, Stream 1 concurrently evaluates the matrix-free
$A_{\text{cart,fn}}$
contribution for interior cells that do not require remote data.
Third, Stream 1 waits for Stream 2 before applying the compatibility
correction where needed. Finally, the solver evaluates the subdomain-boundary part of $A_{\text{cart,fn}}$, the ghost-neighbor contribution $A_{\text{cart,gn}}$, and the ghost-cell interpolation contribution $A_{\text{gcm}}$. This ordering overlaps communication with independent interior work while preserving the data dependencies of the boundary calculations.

\section{Results}

\subsection{Verification}
\begin{figure}[h]
\centering
\includegraphics[width=\textwidth]{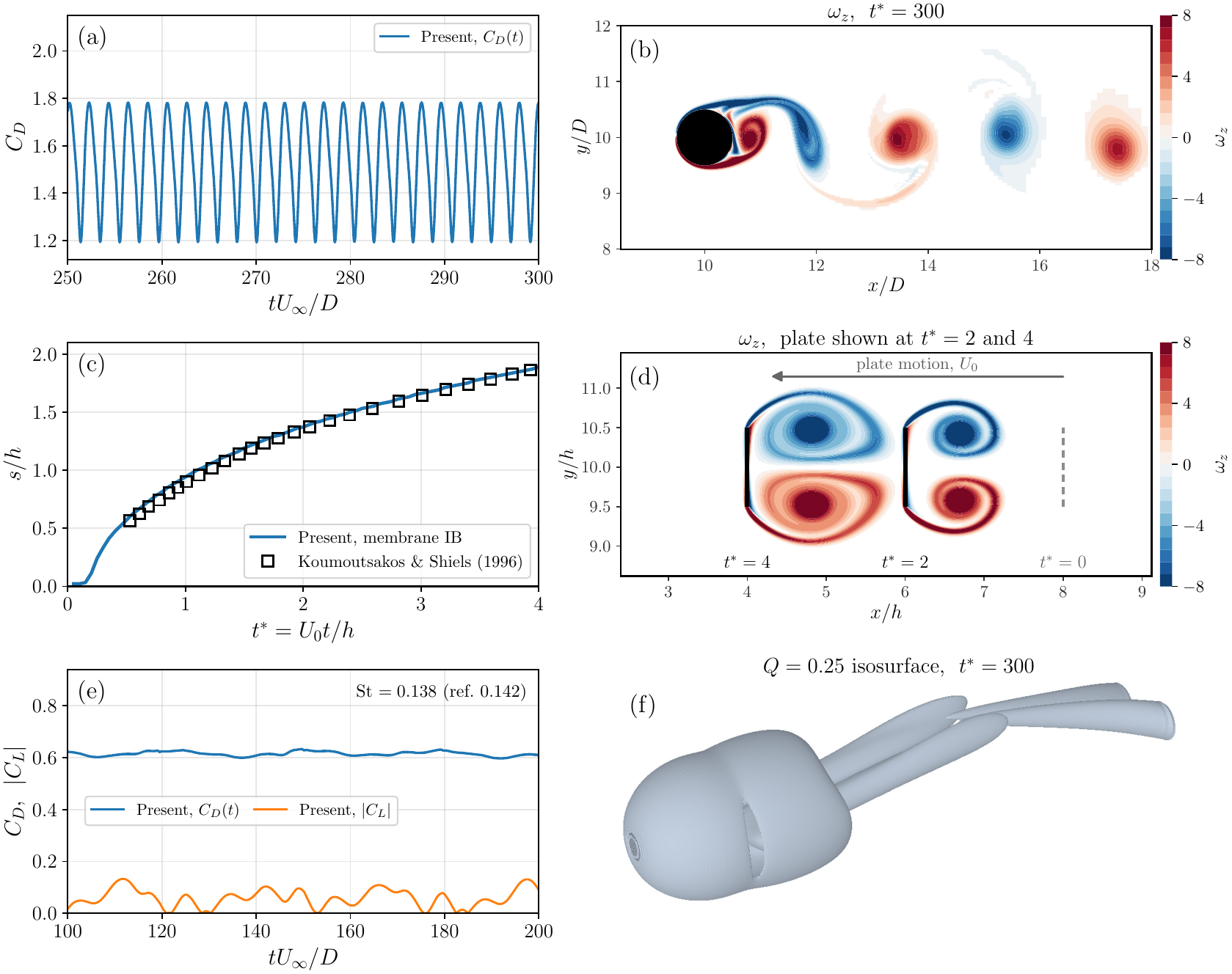}
\caption{Verification cases. (a,b) Cylinder at $Re_D=1000$: $C_D(t)$ and $\omega_z$ at $t^*=300$. (c,d) Impulsively accelerated plate at $Re_h=1000$: $s/h$ and $\omega_z$ at $t^*=2$ and $4$. (e,f) Sphere at $Re_D=350$: $C_D(t)$, $|C_L|(t)$, and a $Q=0.25$ isosurface at $t^*=300$.}
\label{fig:validation}
\end{figure}
Because GPU acceleration requires modifications across nearly all components of the solver, we must verify that the new implementation preserves the accuracy of the original formulation. Three canonical test cases are selected to exercise distinct aspects of the solver. The first case is two-dimensional flow past a stationary circular cylinder. The second is flow generated by a zero-thickness flat plate impulsively accelerated normal to its surface. The third is three-dimensional flow past a stationary sphere. Together, these cases test two- and three-dimensional flows. They also evaluate volumetric and zero-thickness immersed boundaries alongside both stationary and moving-body formulations.

As shown in Figure~\ref{fig:validation}, the accelerated solver reproduces established integral quantities, temporal evolution, and characteristic wake structures for all three cases. For the cylinder, the GPU solution gives $\overline{C}_D=1.53$. This compares with $1.51$ from \cite{henderson1995drag} and $1.45$ from \citep{mittal2008versatile}. The flat-plate recirculation-bubble length has a mean relative difference of $3.0\%$ from the digitized data of \cite{koumoutsakos1996simulations}. For the sphere, the current solver gives $\overline{C}_D=0.62$, compared with $0.64$ reported in \cite{clift2005bubbles}. The corresponding Strouhal number is $St=0.138$, compared with $0.142$ from \cite{mittal2008versatile}. The cylinder and sphere cases verify the conventional sharp-interface treatment for stationary immersed bodies in two and three dimensions. The accelerated flat plate tests the zero-thickness immersed-boundary formulation for a moving boundary. The agreement with published reference data across these complementary cases supports the accuracy of the GPU implementation.

\subsection{Performance Evaluation}
In this section, we describe a comprehensive evaluation of the performance improvements afforded by the GPU-acceleration. 

\subsubsection{Dominant Routines}
Having established the CPU performance bottlenecks in Table~\ref{tab:cpu_component_timing}, we now quantify the speedup achieved by the GPU implementation. The test case is flow past a rectangular wing. Table~\ref{tab:gpu_component_timing} compares the three dominant routines at the same grid resolutions used for the CPU benchmark. The GPU simulations use one, four, and 20 NVIDIA A100 GPUs for the three grids, respectively, while the corresponding CPU simulations use one, four, and ten 24-core Intel Xeon Gold 6248R processor sockets. Each GPU node contains four A100 PCIe GPUs communicating internally via PCIe 4.0 $\times16$ links, which provide a nominal bandwidth of 31.5~GB/s per direction. The 20-GPU simulation spans five nodes and uses HDR100 InfiniBand, providing a nominal bandwidth of 12.5~GB/s shared by the GPUs on each node.

\begin{table}[t]
\centering
\caption{GPU wall time per time step for the principal solver routines in the rectangular-wing case. Parentheses give speedups relative to the CPU implementation.}
\label{tab:gpu_component_timing}
\begin{tabular}{lcccc}
\hline
Grid points
& GPUs
& \texttt{IBlanking} (s)
& AD (s)
& PPE (s) \\
\hline
$256\times320\times320$
& 1
& 0.280 ($2.5\times$)
& 0.039 ($24.4\times$)
& 0.409 ($51.9\times$) \\

$410\times510\times510$
& 4
& 0.286 ($4.4\times$)
& 0.074 ($14.1\times$)
& 1.172 ($18.1\times$) \\

$910\times1110\times1110$
& 20
& 0.624 ($19.6\times$)
& 0.144 ($18.2\times$)
& 2.732 ($20.4\times$) \\
\hline
\end{tabular}
\end{table}

The GPU implementation reduces the cost of all three routines. We define speedups as the CPU wall time divided by the GPU wall time for the same routine and grid, comparing the hardware allocations stated above. With this metric, the largest benefit occurs for the PPE solve, which dominates the CPU runtime and the speedup ranges from approximately 18 to 52. The AD routine also accelerates, showing speedups ranging approximately from 14 to 24. The speedup for \texttt{IBlanking} increases with problem size in these tests, reaching a speedup of nearly 20 on the largest grid. Acceleration therefore extends beyond the regular Cartesian operations to the geometric calculations required by the immersed boundary method. 

\subsubsection{Computational Resource Utilization}
\begin{figure}
\centering
\includegraphics[width=\textwidth]{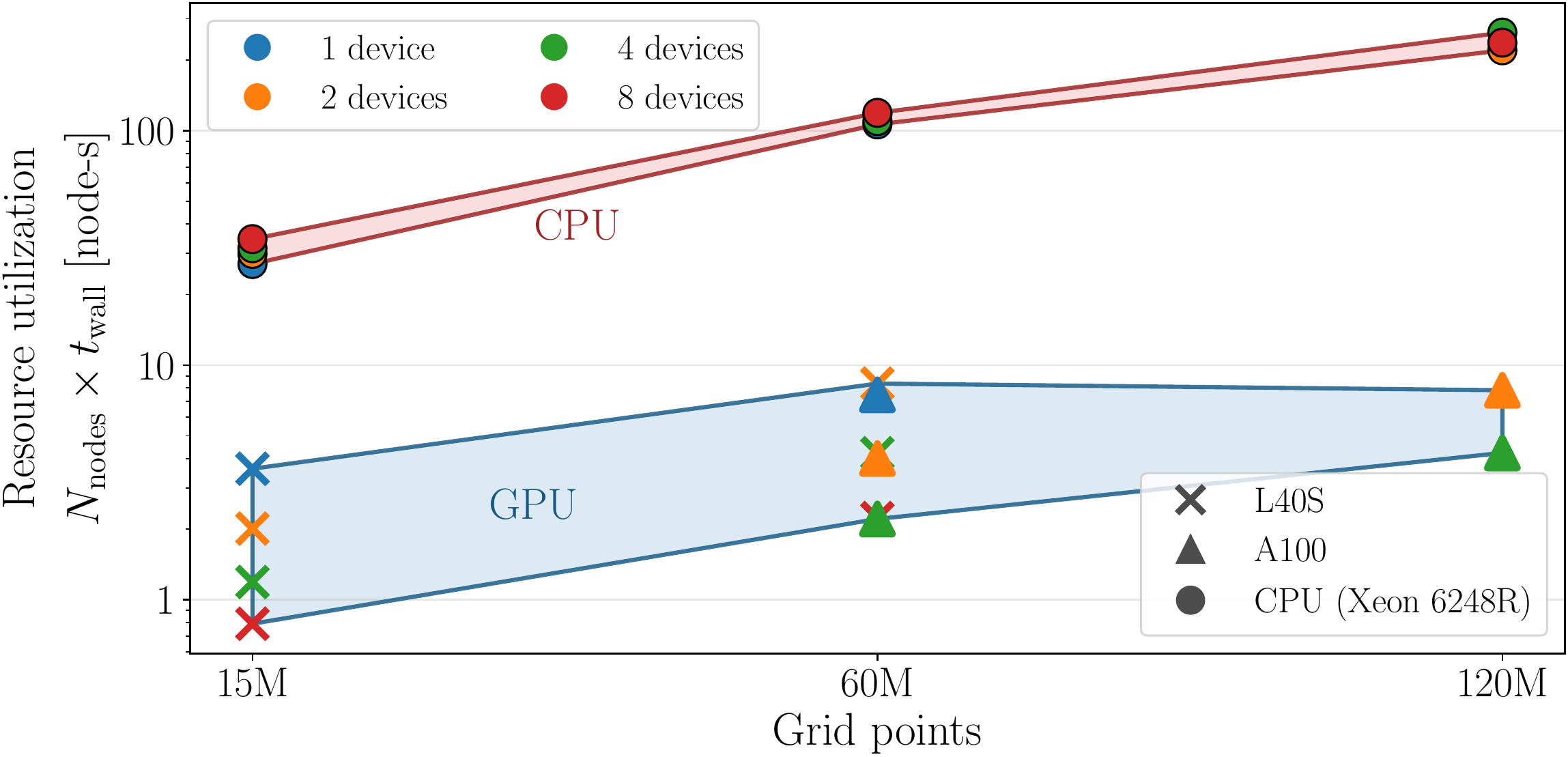}
\caption{Computational resource use, expressed in node-seconds, for the CPU and GPU implementations at three rectangular-wing grid resolutions.}
\label{fig:cpu_vs_gpu}
\end{figure}

Figure~\ref{fig:cpu_vs_gpu} examines what these runtime improvements
mean for computational resource use. The plotted quantity is
node-seconds, defined as the number of allocated compute nodes
multiplied by wall time. These tests use separate grids of
approximately 15, 60, and 120 million points. Each CPU device
represents a complete two-socket Xeon Gold 6248R compute node,
whereas all plotted GPU configurations occupy one compute node
containing either up to four A100 GPUs or up to eight L40S GPUs.
At each grid resolution, the GPU implementation requires fewer
node-seconds than the CPU implementation. Thus, GPU acceleration
reduces both wall-clock runtime and the total allocated node time
needed to complete the calculation. Because all GPU configurations
use one allocated node, any runtime reduction obtained by using
additional GPUs within that node directly reduces node-seconds.
For example, the 60-million-point case requires 106.3 node-seconds
on one CPU node, compared with 2.21 node-seconds using four A100
GPUs on one GPU node, corresponding to a reduction of 97.9\%.

\subsubsection{Strong and Weak Scaling}
We next examine the scaling of the GPU implementation as the computation is
distributed across an increasing number of devices. The tests are performed
on an NVIDIA GB200 NVL72 system, which integrates 72 Blackwell GPUs and 36
Grace CPUs within a single rack. The GPUs reside within a common NVLink
domain, with the fifth-generation interconnect providing a nominal aggregate
bidirectional bandwidth of 1.8~TB/s per GPU. The present tests use allocations
of up to 64 GPUs connected through NVLink. This differs from the A100 PCIe
system considered earlier, where communication beyond a single node traverses
HDR100 InfiniBand. The GB200 tests therefore assess the solver's ability to
exploit a large, tightly coupled multi-GPU domain and identify the
communication costs that remain significant even when high-bandwidth
interconnects are available.

Two three-dimensional flow configurations are used to probe contrasting
workload distributions: flow through a stationary array of 64 spheres and
flow past a shark undergoing prescribed swimming motion
(Fig.~\ref{fig:blanking_schematic}).

The sphere configuration consists of an $8\times8$ array of identical spheres
of diameter $D=1$ in a domain of size $24D\times16D\times8D$. Sphere centers
are separated by $3D$ in the streamwise direction and $2D$ in the transverse
direction, with all spheres located in the plane $z=4D$. A uniform streamwise
velocity of unity is prescribed at the inlet. Zero normal velocity gradients
are imposed at the outlet and the remaining external boundaries, while
homogeneous Neumann conditions are imposed for pressure. The simulations are
performed with a time step of $0.001$. Each sphere is represented
by 13,936 surface vertices and 27,868 triangles, giving a total of 891,904
vertices and 1,783,552 triangles for the full array. This surface
discretization is held fixed as the Cartesian grid is refined. Because the
immersed geometry is uniformly distributed throughout the domain, this case
is designed to provide a well balanced immersed-boundary workload in
addition to an equal Cartesian-cell distribution. At 64 GPUs, each subdomain
contains exactly one sphere.

The shark case contains a single deforming surface represented by 102,905
vertices and 205,805 triangles, as shown in
Fig.~\ref{fig:blanking_schematic}(a). The shark has a nominal body length
$L=1$ and undergoes prescribed lateral deformation according to
\begin{equation}
v_b(\xi,t)
=
-kcA(\xi)\cos\!\left[k(\xi-ct)+\phi\right],
\qquad
A(\xi)=0.02-0.08\xi+0.16\xi^2,
\label{eq:shark_scaling_motion}
\end{equation}
where $\xi=x-x_{\mathrm{ref}}$ is the streamwise coordinate relative to the
motion reference, $k=6.28$, $c=1$, and $\phi=0$. The spanwise surface velocity
is zero. The prescribed velocity is integrated in time to update the body
position, requiring the geometric classification and immersed-boundary
coefficients to be recomputed as the surface moves. Domain dimensions are
expressed in body-length units, and velocities, time steps, and motion
parameters are reported in the solver's non-dimensional units, for which the
prescribed wave speed is unity.

%%%%%%%%%%%%%%%%%%%%%%%%%%%%%%%%%%%%%%

The two cases therefore represent contrasting immersed-boundary workload
distributions. In the sphere array, both the Cartesian and geometric workloads
are distributed approximately uniformly across GPUs. In the shark case, by
contrast, the immersed-boundary work is localized to the subset of partitions
intersected by the single moving body. Together, these configurations allow
the scaling behavior to be examined under both balanced and strongly
localized geometric (unbalanced) workloads.

\begin{figure}[h!]
\centering
\includegraphics[width=\textwidth]{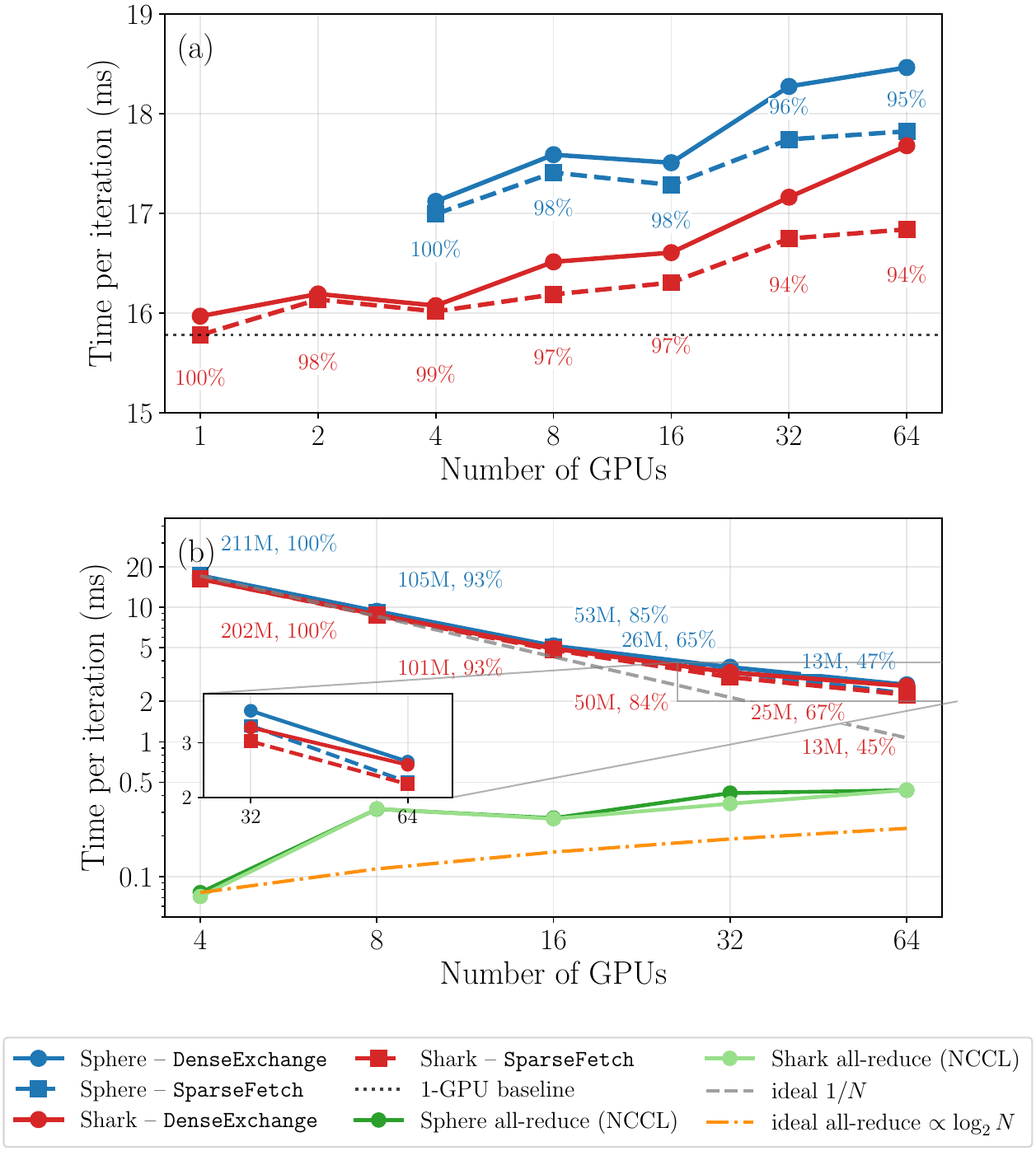}
\caption{Weak scaling (a) and strong scaling (b) of PPE time per iteration for
the sphere-array and moving-shark cases. Green curves show NCCL all-reduce
time. We report scaling efficiency percentage for \texttt{SparseFetch} case
only.}
\label{fig:scaling}
\end{figure}

Figure~\ref{fig:scaling} presents weak scaling in panel (a) and strong scaling
in panel (b). The plotted quantity is the average PPE wall-clock time per
iteration, which isolates the cost of an individual Krylov iteration from
variations in the number of iterations required at different time steps. The
\texttt{DenseExchange} and \texttt{SparseFetch} curves compare the two
ghost-cell communication strategies used by the pressure solver, while the
green curves report the NCCL all-reduce time.

For the weak-scaling tests, the Cartesian grid is increased with GPU count so
that the number of cells per GPU remains approximately constant. The
weak-scaling shark configuration uses a uniform grid over a domain of size
$4\times2\times2$. A uniform streamwise velocity is prescribed at the inlet
and lateral boundaries, a zero-gradient velocity condition is imposed at the
outlet, and homogeneous Neumann pressure conditions are imposed on all
external boundaries. The Reynolds-number parameter is 30,000 and the time
step is $0.0005$. The sphere-array series maintains approximately 211 million
Cartesian cells per GPU, while the shark series maintains approximately
200 million cells per GPU. At 64 GPUs, the sphere calculation reaches a grid
of $3933\times2623\times1313$ points, corresponding to approximately
13.53 billion cells, while the shark calculation reaches
$3715\times1857\times1857$ points, corresponding to approximately
12.79 billion cells. Because the immersed-surface meshes are held fixed, this
construction preserves the Cartesian workload per GPU rather than the surface
workload per GPU. Weak-scaling efficiency is evaluated relative to four GPUs
for the sphere array and one GPU for the shark. The sphere-array setup
requires a minimum of four GPUs because its memory footprint exceeds the
available capacity on one or two GPUs.

Ideal weak scaling corresponds to a constant PPE iteration time as the total
problem size and GPU count increase together. As shown in
Fig.~\ref{fig:scaling}(a), the measured iteration times remain nearly constant
over the full range of GPU counts. With \texttt{SparseFetch}, the sphere-array
and moving-shark cases retain approximately 95\% and 94\% weak-scaling
efficiency, respectively, at 64 GPUs. Thus, the solver sustains nearly constant
PPE throughput per GPU while increasing the total problem size to
approximately 13 billion Cartesian cells. The growing separation between
\texttt{DenseExchange} and \texttt{SparseFetch} also demonstrates the
advantage of communicating only the remote ghost-cell values required by the
local stencil. As the Cartesian problem grows while the surface mesh remains
fixed, ghost-cell data become increasingly sparse relative to the full halo
volume, making dense exchange progressively less efficient.

The strong-scaling tests examine the complementary regime in which a fixed
global problem is distributed over an increasing number of GPUs. The
computational domain has dimensions $15\times20\times20$ and employs a
Cartesian grid stretched in all three coordinate directions. The velocity and
pressure boundary conditions, Reynolds-number parameter, and time step are the
same as those used for the weak-scaling tests. The sphere-array grid contains
$1561\times1041\times521$ points, corresponding to 843.65 million cells,
while the shark grid contains $1701\times951\times501$ points, corresponding
to 807.50 million cells. Both cases are evaluated from four to 64 GPUs.
Because these problems could not be accommodated on a single GPU,
strong-scaling efficiency is measured relative to the four-GPU baseline.

Ideal strong scaling would reduce the PPE iteration time in inverse proportion
to GPU count. Figure~\ref{fig:scaling}(b) shows that both configurations
continue to benefit from increasing GPU count, although their efficiencies
decrease as communication and synchronization consume a larger fraction of
each iteration. With \texttt{SparseFetch}, the strong-scaling efficiencies at
64 GPUs are approximately 47\% for the sphere array and 45\% for the shark.
Despite the markedly different distributions of immersed-boundary work, the
two cases exhibit similar scaling trends. This result demonstrates that the proposed operator decomposition allows the solver to remain less sensitive to unbalanced computational load which exists normally in sharp-interface immersed boundary problems. Moreover, the cost of operator remain similar to matrix-free solves without requiring complex SpMV operations for the dynamics sparse operator.

The benefit of selective ghost-cell communication becomes particularly
evident under strong scaling. As the fixed global grid is divided among more
GPUs, the local computational workload decreases while the amount and latency
of communication become increasingly important. Consequently, the
\texttt{SparseFetch} and \texttt{DenseExchange} curves separate at higher GPU
counts. At 64 GPUs, \texttt{SparseFetch} reduces the total PPE iteration time
by approximately 14\% for both configurations. This reduction is smaller than
the reduction in the measured communication time because a portion of the
exchange is overlapped with computation; nevertheless, the resulting
per-iteration saving accumulates over the many PPE iterations required
throughout a simulation.

At the highest GPU counts, global reductions become the dominant remaining
impediment to strong scaling. Local matrix--vector operations become
progressively shorter as the grid is distributed across more GPUs, whereas
the latency of the NCCL all-reduce operations does not decrease
correspondingly. These reductions therefore occupy an increasing fraction of
the PPE iteration and, unlike the ghost-cell exchanges, cannot be hidden
behind useful Krylov computation. At 64 GPUs, the all-reduce accounts for
approximately 17\% of the PPE iteration time in the \texttt{DenseExchange}
configuration. Because the same NCCL reductions are required for both
communication strategies, \texttt{SparseFetch} reduces ghost-cell
communication cost without eliminating this collective synchronization
overhead. Global reductions therefore emerge as the principal scaling
limitation as the local computational workload decreases on the GB200 system.

\subsection{Capability Demonstration}
Having verified the accuracy, computational performance, and parallel scaling of the GPU-accelerated solver, we next demonstrate its applicability to a range of geometrically and dynamically complex flow problems. The cases are selected to exercise distinct capabilities of the numerical framework, including complex stationary geometries, multi-body particulate flows, and moving-boundary problems. For each case, we describe the computational grid and present three-dimensional flow structures visualized using $Q$-criterion isosurfaces or volume rendering of vorticity, together with relevant quantitative aerodynamic or hydrodynamic measures.

\subsubsection{Flow over an Array of Ellipsoidal Particles at $Re = 10{,}000$}

\begin{figure}[H]
\centering
\includegraphics[width=1\textwidth]{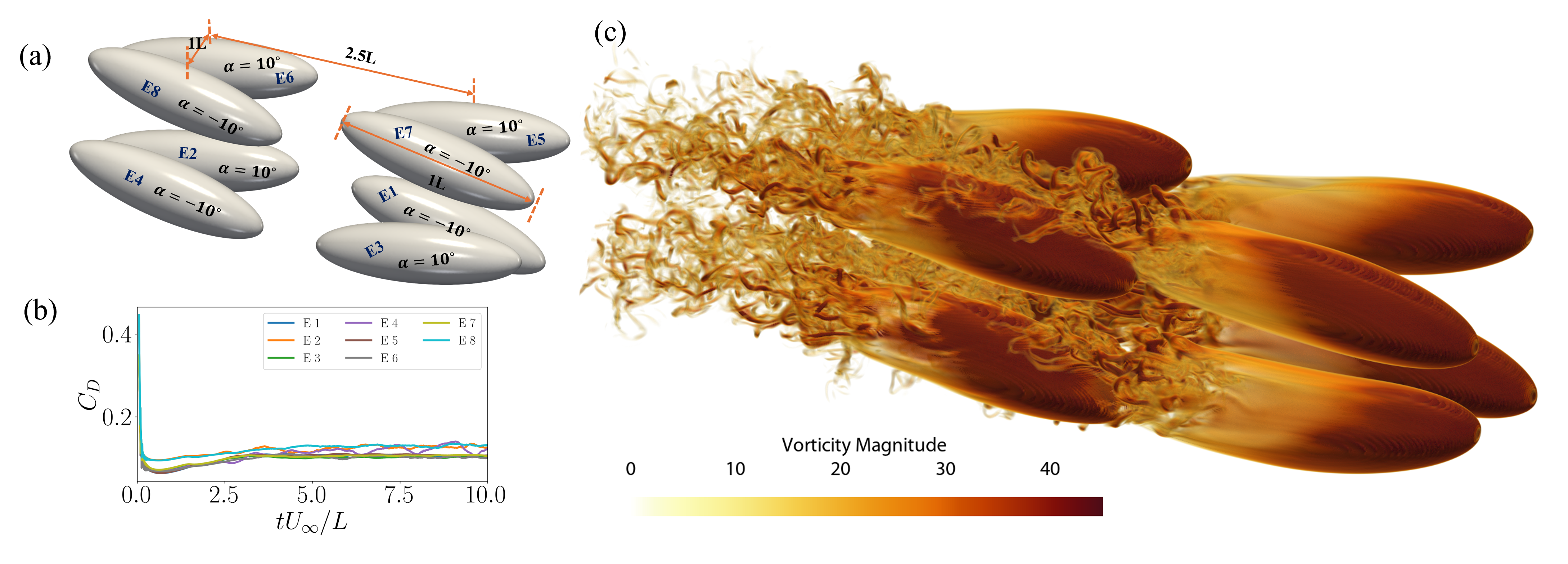}
\caption{Flow through the three-dimensional ellipsoidal array: (a) surface geometry and spatial arrangement of the bodies, (b) instantaneous volumetric rendering of vorticity magnitude, and (c) time histories of the drag coefficient for each individual ellipsoid.}
\label{fig:ell_array}
\end{figure}

Having demonstrated the solver for flow over a single body with complex geometry, we next consider flow through an array of multiple immersed bodies \citep{kumar2026gpu}. The configuration, shown in Fig.~\ref{fig:ell_array}(a), consists of ellipsoids with major-axis length $D$ and a minor-to-major axis ratio of 0.3. The individual ellipsoids are oriented at different angles of attack, resulting in strongly three-dimensional separated flows and interactions among the wakes of neighboring bodies. This configuration therefore provides a useful demonstration of two features of the GPU-accelerated immersed-boundary framework: the ability to accommodate a large number of arbitrarily oriented immersed bodies without body-conformal meshing, and the ability to resolve the complex interacting flow structures generated by a multi-body configuration. Generation of a high-quality body-conformal mesh for such an arrangement would be relatively involved, whereas the Cartesian-grid formulation permits the individual bodies to be incorporated directly through their surface representations.

The characteristic length scale is taken to be the major axis of the ellipsoid, $D$, and the Reynolds number based on $D$ and the freestream velocity $U_\infty$, is $Re=U_\infty D/\nu=10{,}000$. The computational grid contains $1300\times400\times400$ points, corresponding to approximately 208 million grid points. A grid spacing of approximately $0.0025D$ is maintained in the vicinity of the ellipsoids, providing approximately 4--5 grid points across the boundary layers, with the mesh progressively stretched away from the array. The ellipsoid surface mesh contains 61,739 vertices
and 123,474 triangular elements. A non-dimensional time step of $\Delta t=0.0005$ was used.

The simulation was performed on a single compute node using four NVIDIA A100 GPUs connected via PCIe and required approximately 50 hours of wall-clock time to reach $tU_\infty/D=10$. The pressure Poisson solver required, on average, 0.019~s per iteration, including inter-GPU communication, which accounted for approximately 2\% of the iteration time. For comparison, the corresponding pressure-Poisson iteration required 0.41~s on a single Intel Xeon CPU node using two CPU sockets, representing a speedup of approximately 22$\times$ for this portion of the computation. The resulting three-dimensional flow field, visualized using volume rendering of the vorticity magnitude, is shown in Fig.~\ref{fig:ell_array}(b), while the corresponding drag coefficient for each ellipsoid is presented in Fig.~\ref{fig:ell_array}(c).

\subsubsection{Flow over a Sinusoidally Pitching Delta Wing}

\begin{figure}[h!]
\centering
\includegraphics[width=1\textwidth]{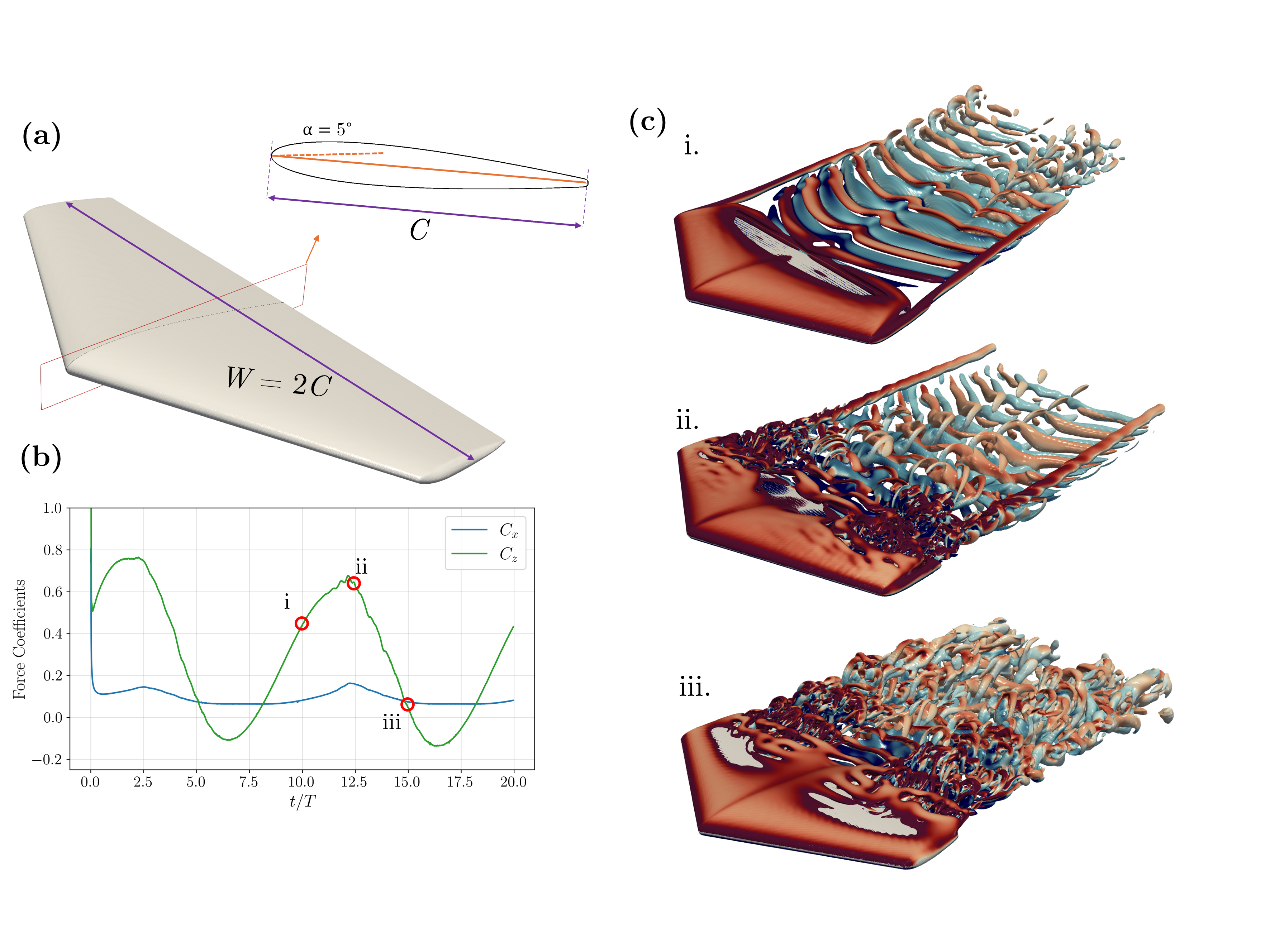}
\caption{Pitching delta wing with NACA 0015 cross-section undergoing a
half-sine pitching maneuver.(a) Wing geometry. (b) Lift and drag histories.
(c) Instantaneous $Q$-criterion isosurfaces colored by spanwise velocity.}
\label{fig:swept_wing}
\end{figure}

The preceding cases demonstrate the performance of the GPU-accelerated solver for stationary immersed bodies. We next consider a moving-boundary problem involving a pitching delta wing. In contrast to the stationary-body cases, motion of the immersed surface requires the ghost-cell classification and associated GCM interpolation stencils to be updated at every time step, thereby introducing additional computational complexity. The delta wing has a NACA 0015 cross-section and a leading-edge sweep angle of $\Lambda=27^\circ$. The midspan chord length, $c$, is used as the characteristic length scale, as shown in Fig.~\ref{fig:swept_wing}(a).

The wing undergoes a prescribed pitching maneuver in which the angle of attack varies according to a half-sine profile,
\begin{equation}
\alpha(t) = \alpha_{\max}
\sin\left(\frac{\pi t}{T_p}\right),
\qquad 0 \leq t \leq T_p,
\end{equation}
where $\alpha_{\max}=5^\circ$ is the maximum angle of attack and
$T_p=5$ is the duration of the half-sine pitching maneuver
in non-dimensional time units. The wing therefore pitches from $\alpha=0^\circ$ to $\alpha_{\max}$ and subsequently returns to $\alpha=0^\circ$. This prescribed motion generates an unsteady leading-edge vortex (LEV) during the pitch-up phase, followed by its downstream convection and shedding during pitch-down. The case thus provides a useful demonstration of the dynamic ghost-cell blanking and GCM interpolation updates required for moving immersed boundaries.

The Reynolds number, based on the freestream velocity $U_\infty$ and midspan chord $c$, is $Re=U_\infty c/\nu=10{,}000$. The computational domain is discretized using a Cartesian grid containing $901\times1081\times301$ points, corresponding to approximately 293 million grid points, with increased resolution in the vicinity of the wing to resolve the boundary layer and near-wake flow. The surface mesh contains 133,688 vertices
and 267,372 triangular elements. The simulation was performed using four NVIDIA RTX PRO 6000 Blackwell GPUs. A nondimensional time step of $\Delta t=0.0005$ was employed, resulting in a maximum CFL number of approximately 0.35. The instantaneous flow structures shown in Fig.~\ref{fig:swept_wing}(c)
illustrate the formation and downstream evolution of the LEV over the
suction surface. The corresponding histories of the lift and drag
coefficients are presented in Fig.~\ref{fig:swept_wing}
(b), with the lift exhibiting a pronounced maximum
associated with the development of the LEV, followed by a rapid decrease
as the vortex convects downstream and sheds from the wing. Over the sampled time steps, the average PPE wall
time is 0.0185~s per iteration, while the \texttt{IBlanking}
pass requires an average of 2.438~s per time step.

\subsubsection{Aero-Structural Dynamics of Bat Flight}
The preceding cases demonstrate the GPU-accelerated framework for stationary and prescribed moving boundaries. Here, we consider the fully coupled fluid--structure interaction associated with a bat-inspired flexible membrane wing in forward flight. This problem combines complex three-dimensional flapping kinematics, large membrane deformations, and strongly unsteady vortical flow. The wing morphology is based on \emph{Pteropus pumilus} and comprises the propatagium, plagiopatagium, dactylopatagium major, and dactylopatagium medius, with the skeletal joint kinematics prescribed from high-speed videogrammetry \citep{riskin2008quantifying}. The flow is solved using the GPU-accelerated \texttt{ViCar3D} while the deformation of the flexible wing membrane is determined through coupling with a structural model.

The membrane is represented by a zero-thickness spring-network model whose dynamics are governed by
\begin{equation}
M\frac{d^2\mathbf{X}}{dt^2}
=
\mathbf{F}_{\mathrm{ext}}
-\mathbf{F}_{\mathrm{int}}
-\zeta\frac{d\mathbf{X}}{dt},
\end{equation}
where $\mathbf{X}$ denotes the membrane-node position, $M$ is the corresponding nodal mass, $\mathbf{F}_{\mathrm{ext}}$ represents the aerodynamic loading, $\mathbf{F}_{\mathrm{int}}$ accounts for the internal elastic and bending forces, and $\zeta$ is a structural damping coefficient. The aerodynamic loads computed by the flow solver drive the membrane deformation, which in turn modifies the immersed boundary seen by the fluid solver. Owing to the relatively large solid-to-fluid density ratio for flight in air, the fluid and structural solvers are coupled explicitly at each time step.
\begin{figure}[h!]
\centering
\includegraphics[width=\textwidth]{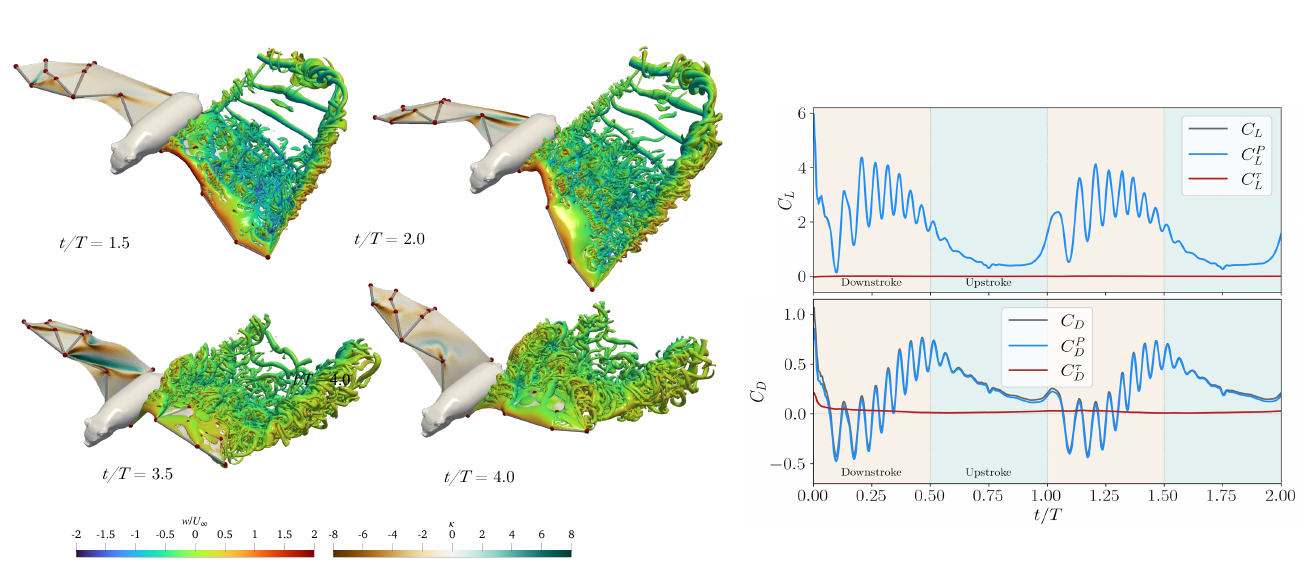}
\caption{
Bat-wing fluid--structure interaction.
Left: instantaneous wake structures and membrane deformation.
Right: lift and drag histories.}
\label{fig:bat_5k}
\end{figure}

The simulation is performed at $Re=5{,}000$ using a Cartesian grid containing $750\times754\times800$ points, corresponding to approximately 452 million grid points. The grid is refined in the vicinity of the wing and its wake to resolve the boundary layers and the multiscale vortical structures generated during the flapping cycle. The simulation was performed using four RTX Blackwell Pro 6000 GPUs with a non-dimensional time step of $\Delta t=0.0005$.

Representative instantaneous flow structures and the corresponding membrane
deformation over the flapping cycle are shown in Fig.~\ref{fig:bat_5k}.
The snapshots are shown at $t/T=1.5$, $2.0$, $3.5$, and $4.0$.
The wake is colored by vertical velocity $w/U_\infty$, and the membrane is
colored by curvature $\kappa$.
During the downstroke, strong leading-edge vortices (LEVs) develop over the
wing, with the LEV dynamics varying substantially across the different
membrane panels. In particular, vortex formation and shedding occur over the propatagium, while a more persistent attached LEV develops over portions of the dactylopatagium medius and major. The aerodynamic loading associated with these vortical structures produces substantial spatial and temporal variations in membrane deformation and curvature.

The corresponding histories of the lift ($C_L$) and drag ($C_D$) coefficients exhibit pronounced high-frequency fluctuations during the downstroke. These fluctuations are associated with the coupled response of the flexible membrane and the unsteady formation and shedding of vortical structures \citep{kumar2025computational}. In comparison, the upstroke produces smoother aerodynamic force histories while maintaining a reduced but positive lift \citep{kumar2025computational}. For the bat timing configuration, the average PPE
wall time is 0.0352~s per iteration, while the \texttt{IBlanking}
pass requires an average of 0.477~s per time step. This example therefore demonstrates the ability of the GPU-accelerated framework to simulate large-scale, fully coupled fluid--structure interaction involving complex moving geometries and highly unsteady three-dimensional flows.

\section{Conclusions}

This work presents a GPU-native, multi-GPU implementation of the sharp-interface immersed boundary solver \texttt{ViCar3D} for large-scale incompressible flow simulations involving complex moving and deforming boundaries. Such problems pose challenges that extend beyond accelerating the conventional structured-grid flow solver: body motion continuously changes the fluid--solid topology, ghost-cell interpolation stencils, and immersed-boundary contributions to the pressure operator, while introducing sparse and spatially nonuniform communication. The approach developed here exploits the regularity of the underlying Cartesian discretization while isolating these dynamically evolving geometric and algebraic contributions.

Several complementary algorithmic developments enable efficient execution on GPUs. Geometric classification is accelerated by restricting expensive surface searches to a compact corona surrounding the immersed geometry and by fusing geometric operations to reduce device memory traffic. Small interpolation systems associated with ghost and fresh cells are processed using batched GPU linear algebra. For the pressure Poisson equation, the dynamically varying operator is decomposed into matrix-free Cartesian fluid-neighbor contributions, sparse ghost-neighbor corrections, and compact ELLPACK-formatted ghost-cell equations. This decomposition avoids repeated construction of a monolithic sparse matrix and permits the dominant Cartesian operations to retain regular, coalesced memory access. A Jacobi-preconditioned BiCGStab solver further combines CUDA-graph execution, fused reductions, dual-stream computation--communication overlap, and NVSHMEM-based selective retrieval of remote ghost-cell data.

These developments produce substantial improvements over the legacy MPI-based CPU implementation. Across the rectangular-wing benchmarks, the GPU implementation accelerates the pressure Poisson solve by approximately $18$--$52\times$, the advection--diffusion solve by $14$--$24\times$, and the geometric classification operation by up to approximately $20\times$. The reduction in wall-clock time is accompanied by a substantial decrease in computational resource usage; for example, the 60-million-point case requires 97.9\% fewer node-seconds using four A100 GPUs than using a single two-socket CPU node.

The multi-GPU results demonstrate that these gains can be sustained at much larger scales. On the GB200 NVL72 system, \texttt{SparseFetch} retains approximately 95\% and 94\% weak-scaling efficiency at 64 GPUs for the sphere-array and moving-shark configurations, respectively, with problem sizes approaching 13 billion Cartesian cells. Strong-scaling efficiencies at 64 GPUs are approximately 47\% and 45\%, despite the markedly different distributions of immersed-boundary work in the two cases. Selective ghost-cell communication reduces PPE iteration time by approximately 14\% at 64 GPUs relative to dense exchange. At high GPU counts, the dominant scaling limitation shifts from sparse boundary communication to the global reductions required by BiCGStab and the pressure-compatibility constraint, indicating that collective synchronization rather than local immersed-boundary processing ultimately limits strong scaling.

The capability demonstrations span multi-body flow through an
ellipsoidal-particle array, prescribed motion of a pitching delta wing, and
fully coupled aero-structural dynamics of a deforming bat wing. Together,
these cases demonstrate that the framework preserves the geometric flexibility
of the sharp-interface formulation while accommodating volumetric and
zero-thickness bodies, moving and deforming surfaces, and coupled
fluid--structure interaction over a wide range of problem sizes. More broadly,
the strategies developed here including localization of dynamic geometry,
decomposition of static and evolving operators, sparse communication, and
asynchronous execution, address computational challenges common to
moving-boundary CFD and computational mechanics.

The principal remaining limitation is the strong-scaling ceiling imposed by
global synchronization and communication latency. At local problem sizes below
approximately 50 million grid points per GPU, communication increasingly
dominates the iteration cost. Future improvements can include dynamic load balancing for spatially nonuniform workloads
arising from translating and deforming immersed bodies. Moreover, the newer generation of GPUs are increasingly moving towards low precision computing. Here, mixed precision or emulation strategies can extend the scale and efficiency of GPU-based sharp-interface
immersed boundary simulations and other dynamically evolving sparse systems.

\section*{Acknowledgment}
We acknowledge support from the Office of Naval Research (N000142412516, N000142612289, and N000142612365), the National Science Foundation (CBET-2436107 and CBET-2618840), the Army Research Office (W911NF261A095), and the NVIDIA Academic Grant Program. We also acknowledge computational resources provided by Johns Hopkins University’s Advanced Research Computing at Hopkins (ARCH) and the Department of Defense High Performance Computing Modernization Program.

\appendix

\section{Normal Projection and Barycentric Coordinates}
\label{app:barycentric}
A triangle $\tau=(m_1,m_2,m_3)$ is specified by its three vertex
indices, with corresponding coordinates $\mathbf{x}_{m_1}$,
$\mathbf{x}_{m_2}$, and $\mathbf{x}_{m_3}$. These correspond to
vertices $A$, $B$, and $C$ in Figure~\ref{fig:blanking_schematic}(e).
For a Cartesian point $\mathbf{x}_q$ and triangle unit normal
$\hat{\mathbf{n}}_\tau$, the normal projection onto the triangle
plane is
\begin{equation}
\mathbf{x}_{\mathrm{BI}}^{*}
=
\mathbf{x}_q
-
\left[
\hat{\mathbf{n}}_\tau\cdot
(\mathbf{x}_q-\mathbf{x}_{m_1})
\right]\hat{\mathbf{n}}_\tau.
\label{eq:app_normal_projection}
\end{equation}
Define the triangle-edge vectors and projected displacement as
\begin{equation}
\mathbf{u}=\mathbf{x}_{m_2}-\mathbf{x}_{m_1},
\qquad
\mathbf{v}=\mathbf{x}_{m_3}-\mathbf{x}_{m_1},
\qquad
\mathbf{w}=\mathbf{x}_{\mathrm{BI}}^{*}-\mathbf{x}_{m_1}.
\end{equation}
The barycentric coordinates satisfy
$\mathbf{w}=S\mathbf{u}+T\mathbf{v}$.
For a nondegenerate triangle, let
$D=(\mathbf{u}\cdot\mathbf{u})(\mathbf{v}\cdot\mathbf{v})
-(\mathbf{u}\cdot\mathbf{v})^2>0$.
Then
\begin{equation}
\begin{aligned}
S &=
\frac{
(\mathbf{v}\cdot\mathbf{v})(\mathbf{w}\cdot\mathbf{u})
-
(\mathbf{u}\cdot\mathbf{v})(\mathbf{w}\cdot\mathbf{v})
}{D},
\\
T &=
\frac{
(\mathbf{u}\cdot\mathbf{u})(\mathbf{w}\cdot\mathbf{v})
-
(\mathbf{u}\cdot\mathbf{v})(\mathbf{w}\cdot\mathbf{u})
}{D}.
\end{aligned}
\label{eq:app_barycentric}
\end{equation}
The projected point lies within or on the edges of the triangle when
\begin{equation}
S\geq0,
\qquad
T\geq0,
\qquad
S+T\leq1.
\end{equation}
A projection satisfying these conditions is retained as a candidate
boundary intercept.

\section{Implementation of Immersed-Boundary Matrix Operators}
\label{app:matrix_operators}

This appendix describes the pressure-operator contributions introduced
in Section~\ref{sec:solver}:
\begin{equation}
A\mathbf{x}
=
A_{\mathrm{cart,fn}}\mathbf{x}
+
A_{\mathrm{cart,gn}}\mathbf{x}
+
A_{\mathrm{gcm}}\mathbf{x}.
\end{equation}
The combined vector $\mathbf{x}$ contains the regular pressure field
$p$ and the separate ghost pressure field $p_G$. The Cartesian
contributions act on distinct sets of fluid-cell rows.
$A_{\mathrm{cart,fn}}$ applies to cells with only fluid neighbors,
while $A_{\mathrm{cart,gn}}$ applies to cells with ghost neighbors.
$A_{\mathrm{gcm}}$ supplies the ghost-cell equations. Prescribed
boundary data and compatibility corrections are handled separately.

\subsection{Cartesian and Ghost-Neighbor Operators}

Let $\mathbf{i}=(i,j,k)$ denote a fluid-cell index and $\mathcal{D}$
the six Cartesian neighbor offsets. The central and neighbor
coefficients are denoted by $a_{C,\mathbf{i}}$ and
$a_{\mathbf{d},\mathbf{i}}$, respectively.

For cells with only fluid neighbors, the matrix-free seven-point
stencil gives
\begin{equation}
[A_{\mathrm{cart,fn}}\mathbf{x}]_{\mathbf{i}}
=
a_{C,\mathbf{i}}p_{\mathbf{i}}
+
\sum_{\mathbf{d}\in\mathcal{D}}
a_{\mathbf{d},\mathbf{i}}p_{\mathbf{i}+\mathbf{d}}.
\end{equation}

For ghost-neighbor cells, the binary mask
$\delta_{\mathbf{i},\mathbf{d}}$ determines whether a neighboring
contribution is obtained from $p$ or $p_G$. The mask is zero for
the regular field and one for the ghost field. The corresponding
row is
\begin{equation}
\begin{aligned}
[A_{\mathrm{cart,gn}}\mathbf{x}]_{\mathbf{i}}
&=a_{C,\mathbf{i}}p_{\mathbf{i}}\\
&\quad+
\sum_{\mathbf{d}\in\mathcal{D}}
a_{\mathbf{d},\mathbf{i}}
\Bigl[
(1-\delta_{\mathbf{i},\mathbf{d}})
p_{\mathbf{i}+\mathbf{d}}
+
\delta_{\mathbf{i},\mathbf{d}}
(p_G)_{\mathbf{i}+\mathbf{d}}
\Bigr].
\end{aligned}
\label{eq:nb_blend}
\end{equation}
Only ghost-neighbor rows require this field selection, avoiding
unnecessary ghost-field accesses throughout the Cartesian interior.

\subsection{Ghost-Cell Operator}

For each of the $N_{\mathrm{GC}}$ active ghost cells, the GCM
interpolation uses eight coefficients $C(r,n)=\beta_{r,n}$,
where $r=1,\ldots,8$ and $n=1,\ldots,N_{\mathrm{GC}}$.
These coefficients are the Neumann image-point interpolation weights.

Let $\mathbf{g}_n=(i_n,j_n,k_n)$ denote the ghost-cell index and
$\mathbf{i}_{0,n}$ the base index of its interpolation cell.
The stencil-node indices are
\begin{equation}
\mathbf{s}_{r,n}
=
\mathbf{i}_{0,n}
+
(\Delta i_r,\Delta j_r,\Delta k_r),
\label{eq:app_stencil_indices}
\end{equation}
where the offsets identify the eight corners of the interpolation
cell.
Define
\begin{equation}
\mathcal{R}_n
=
\{r:\mathbf{s}_{r,n}\neq\mathbf{g}_n\}.
\end{equation}
If the target ghost cell is a stencil corner, its row in the local
interpolation matrix is replaced by the prescribed normal-derivative
condition at the boundary intercept. This slot contributes boundary
data to the right-hand side and is excluded from the operator sum.

The equation associated with ghost cell $n$, located at
$\mathbf{g}_n$, is
\begin{equation}
\begin{aligned}
[A_{\mathrm{gcm}}\mathbf{x}]_n
&=(p_G)_{\mathbf{g}_n}\\
&\quad-
\sum_{r\in\mathcal{R}_n}
C(r,n)
\Bigl[
(1-\sigma_{r,n})p_{\mathbf{s}_{r,n}}
+
\sigma_{r,n}(p_G)_{\mathbf{s}_{r,n}}
\Bigr],
\end{aligned}
\label{eq:gcm_row}
\end{equation}
where $\sigma_{r,n}\in\{0,1\}$ selects the regular or ghost field.

Let $\chi_G(\mathbf{s})$ equal one at an active ghost cell and zero
otherwise. For a membrane, let $\epsilon_n(\mathbf{s})\in\{-1,1\}$
denote the side classification of point $\mathbf{s}$ relative to
the membrane associated with ghost cell $n$. Then
\begin{equation}
\sigma_{r,n}
=
\begin{cases}
\chi_G(\mathbf{s}_{r,n}),
& \text{volumetric solid},\\
\chi_G(\mathbf{s}_{r,n})
\mathbf{1}\!\left[
\epsilon_n(\mathbf{s}_{r,n})
=
\epsilon_n(\mathbf{g}_n)
\right],
& \text{membrane}.
\end{cases}
\end{equation}
Thus, a stencil node on the opposite side of a membrane contributes
through the regular field. The leading ghost-cell unknown has unit
coefficient and is applied separately from the interpolation terms.

\subsection{GPU Storage and Geometry Updates}

The ghost-cell interpolation data use fixed-width ELLPACK storage
with at most eight coefficients per row~\citep{ZHENG20142639}.
One GPU thread evaluates each row. The coefficient array has
dimensions $(8,N_{\mathrm{GC}})$, with the eight coefficients of
each row contiguous under Fortran column-major storage.
Stencil indices are reconstructed from the interpolation-cell
base indices and shared corner offsets.

Body motion updates the active ghost list, interpolation-cell bases,
coefficients, and ghost-neighbor masks and lists. The regular
Cartesian stencil retains its fixed connectivity and matrix-free
evaluation. The interpolation data therefore require
$8N_{\mathrm{GC}}$ coefficients and
$\mathcal{O}(N_{\mathrm{GC}})$ geometric indices without assembling
the full pressure matrix.

\bibliographystyle{elsarticle-harv} 
\bibliography{ref}

\end{document}